\documentclass[aip,reprint,jcp]{revtex4-1} % for emulating the final form
\usepackage[%
  colorlinks=true,
  urlcolor=blue,
  linkcolor=blue,
  citecolor=black
]{hyperref}
\usepackage{graphicx}
\usepackage{dcolumn}
\usepackage{bm}
\usepackage[utf8]{inputenc}
\usepackage[T1]{fontenc}
\usepackage{mathptmx}
\usepackage{etoolbox}
\usepackage{amsmath}
\renewcommand{\selectlanguage}[1]{}
\DeclareUnicodeCharacter{2009}{\,} 
\DeclareUnicodeCharacter{2032}{'}

\makeatletter
\def\@email#1#2{%
 \endgroup
 \patchcmd{\titleblock@produce}
  {\frontmatter@RRAPformat}
  {\frontmatter@RRAPformat{\produce@RRAP{*#1\href{mailto:#2}{#2}}}\frontmatter@RRAPformat}
  {}{}
}%
\makeatother

\begin{document}
\preprint{AIP/123-QED}

\def\*#1{\mathbf{#1}}
\def\&#1{#1}

\title{Stockmayer Fluid with a Shifted Dipole: Interfacial Behavior}

\author{Samuel Varner}
\altaffiliation{These authors contributed equally.}
\affiliation{Division of Chemistry and Chemical Engineering, California Institute of Technology, Pasadena, CA 91125, United States}

\author{Pierre J. Walker}
\altaffiliation{These authors contributed equally.}
\affiliation{Division of Chemistry and Chemical Engineering, California Institute of Technology, Pasadena, CA 91125, United States}

\author{Ananya Venkatachalam}
\altaffiliation{These authors contributed equally.}
\affiliation{Department of Chemistry, Harvey Mudd College, Claremont, CA 91711, United States}

\author{Bilin Zhuang}
\affiliation{Department of Chemistry, Harvey Mudd College, Claremont, CA 91711, United States}

\author{Zhen-Gang Wang}
\affiliation{Division of Chemistry and Chemical Engineering, California Institute of Technology, Pasadena, CA 91125, United States}
\email{zgw@caltech.edu}

\date{\today}

\begin{abstract}
We investigate the properties of the liquid--vapor interface in the shifted Stockmayer fluid using molecular dynamics simulations in the canonical ensemble. We study the role of the dipole moment strength and the degree of asymmmetry on equilibrium interfacial characteristics, including density profiles, polar order, nematic order, interfacial polarization, electric field, and electrostatic potential. In addition, we compute angular distribution functions across the interface to gain insight into how the dipole shift affects the molecular orientation. We find that the shift significantly effects angular distribution functions by altering the polar order while leaving the nematic order relatively unaffected, in comparison to the reference symmetric Stockmayer fluid. We find that these results are consistently explained using an image-dipole construction that has been previously applied to symmetric Stockmayer fluids but has never been extended to the shifted model. We find remarkable agreement between the simple theory and the simulations in the qualitative shape of the distribution functions for both the liquid and vapor phases in proximity to the interface. Unexpectedly, the spontaneous polarization at the interface, and therefore the generated electric field, changes sign as the dipole moment strength increases. This also leads to an inversion of the sign of the potential difference across the interface.
\end{abstract}

\maketitle

%%%%%%%%%%%%%%%%%%%%%%%%%%%%%%%%%%%%%%%%%%%%%%%%%%%%%%%%%%%%%%%%%%%%%
%% Start the main part of the manuscript here.
%%%%%%%%%%%%%%%%%%%%%%%%%%%%%%%%%%%%%%%%%%%%%%%%%%%%%%%%%%%%%%%%%%%%%
\section{Introduction}

It has been well documented that reaction rates can increase by several orders of magnitude in systems with confined volumes and large interfacial area, such as microemulsions.\cite{yan_organic_2016,wei_accelerated_2020,narayan_water_2005,serrano-luginbuhl_soft_2018,jordan_photooxidation_2022,chen_spontaneous_2023,liu_oxidation_2021,wan_mechanistic_2023,rossignol_atmospheric_2016} Interfaces between water and other phases, including air and oil, are known to promote certain reactions through a process often referred to as “on-water” catalysis.\cite{narayan_water_2005} This phenomenon is especially important in atmospheric chemistry, where abundant microdroplets can catalyze many reactions of atmospheric relevance.\cite{xia_counterintuitive_2023,liang_water_2023,george_heterogeneous_2015} Despite their significance, the molecular mechanisms responsible for this unique form of catalysis remain elusive and actively debated.\cite{ruiz-lopez_molecular_2020} Explanations for the enhanced reactivity at water interfaces include evaporation-induced reactant enrichment,\cite{yan_chemical_2013} pH changes,\cite{girod_accelerated_2011} partial solvation environments,\cite{wei_accelerated_2020,pestana_dielsalder_2020} spatially varying dielectric properties,\cite{matyushov_electrostatic_2019} and orientational changes near the interface driven by entropic or enthalpic effects.\cite{munoz-santiburcio_confinement-controlled_2021}

Perhaps the most compelling physical phenomenon that has been thought to drive interfacial chemistry is the presence of a strong electric field. Namely, if water (or another polar solvent) shows a preferential orientation at the interface, then an electric field will be spontaneously generated that can potentially stretch reactive bonds and lower the activation barrier.\cite{welborn_computational_2018,ashton_ab_2020,shaik_electric-field_2020} In fact, enzymes catalyze reactions in a similar manner, by orienting reactant molecules such that reactive bonds align with strong directional electric fields inside the active site.\cite{bim_local_2021,eberhart_electric_2024,aragones_electrostatic_2016,shaik_oriented_2016,stuyver_external_2020,zheng_two-directional_2022,welborn_computational_2018,welborn_fluctuations_2019} In the case of the air--water interface, vibrational sum frequency generation (vSFG) measurements have revealed that water preferentially orients with a "dangling" O-H bond at the surface,\cite{morita_theoretical_2000,du_vibrational_1993,medders_dissecting_2016,moberg_temperature_2018} supported underneath by a 2-dimensional hydrogen bond network.\cite{pezzotti_2d_2017,pezzotti_2d-hb-network_2018} These dangling O-H bonds can serve as probes for the local electric field. The Stark shifts measured in Infrared Photodissociation (IRPD) spectroscopy have revealed a significant electric field at the air--water interface.\cite{cooper_structural_2017} Recently the electric field at an air-oil interface was quantified via Stimulated Raman excited flourescence (SREF) microscopy, where a magnitude of $10$MV/cm was reported. This is also in agreement with electric fields recently computed from ReaxFF/C-GeM molecular dynamics (MD) simulations,\cite{hao_can_2022} while ab initio calculations predicted much stronger fields on the order of hundreds of MV/cm.\cite{kathmann_understanding_2011} Nevertheless, these results all indicate the presence of electric fields strong enough to lower activation barriers or drive bond breaking.

For good reason, most studies of interfacial electric fields have focused on air--water and oil--water systems because of their atmospheric and biological relevance. Water has many distinctive features, including high polarity, autoionization, and a strong hydrogen bonding network.\cite{stillinger_water_1980,franks_water_2000,ball_lifes_2001} The combination of these effects makes it difficult to determine the true origin of the strong electric fields observed at the water interface. In addition, results obtained specifically for water cannot be easily applied to other liquids that do not share its special bulk properties. To better identify the general physical factors that can generate interfacial polarization, coarse-grained models are especially useful.

The Stockmayer fluid\cite{stockmayer_second_1941,stockmayer_second_1941-1} (SF) is by far the most studied model for general polar fluids. In this model, molecules are represented as Lennard-Jones spheres with a point dipole located at the center. A large body of work has explored the properties of the SF model and its connections to real fluids. Research has examined ion solvation energy and dynamics,\cite{shock_solvation_2020,bagchi_solvation_2010,perera_dynamics_1992} dielectric properties,\cite{adams_static_1981,pollock_static_1980,shock_molecular_2023} ferroelectric transitions,\cite{groh_ferroelectric_1994,weis_ferroelectric_1993,pounds_are_2007,bartke_dielectric_2006} and phase equilibria.\cite{marx_phase_2022,marx_vapor-liquid_2023,moore_liquid-vapor_2015} The liquid--vapor interface of the SF model has also been the subject of extensive theoretical and simulation studies. These works show that spherical particles with permanent dipoles adopt preferential orientations across the interface that are largely determined by strong dipolar interactions. Classical density functional theory (cDFT) and integral equation theory predict that dipoles align parallel to the interface on the liquid side and perpendicular on the vapor side.\cite{frodl_bulk_1992,frodl_thermal_1993,iatsevitch_structure_2000} MD simulations of the SF model support the parallel orientation within the liquid but generally do not reproduce the perpendicular orientation in the vapor.\cite{moore_liquid-vapor_2015,mecke_molecular_2001,enders_molecular_2004,eggebrecht_liquidvapor_1987}However, because of the spherical symmetry of the molecules, the SF model cannot produce a net electric field at the interface (in the absence of a ferroelectric transition).

We hypothesize that introducing slight molecular asymmetry into the SF model is sufficient to break the symmetry at the interface and generate an electric field. In real molecules, the geometric structure does not always coincide with the charge distribution, which can shift the effective dipole toward one side of the molecule. The simplest way to capture this effect is with a spherical particle that carries a point dipole displaced by a distance $d$ from its geometric center. This extension of the SF model is known as the shifted Stockmayer fluid (sSF).\cite{langenbach_co-oriented_2017} Although still highly coarse-grained, the sSF model is more realistic than the SF model, since nearly all real dipolar molecules have non-centered dipoles. The earliest use of the sSF model we found in the literature is by Kusaka et al., who applied it to study the sign effect in ion-induced water-droplet nucleation in the atmosphere.\cite{kusaka_ion-induced_1995} Subsequently, the model received little attention until Langenbach introduced the Co-Oriented Fluid Functional Equation for Electrostatic interactions (COFFEE) in 2017.\cite{langenbach_co-oriented_2017} This perturbative approach enables accurate predictions of the properties of homogeneous liquids and vapors for both SF and sSF models at a fraction of the cost of molecular dynamics simulations, though it is restricted to certain dipole strengths and shifts. However, because it is formulated for homogeneous fluids, COFFEE cannot be applied directly to interfacial properties such as density and electric field profiles. To complement and validate this theory, Marx, Kohns, and Langenbach have performed extensive MD and Monte Carlo simulations to characterize the liquid--vapor equilibrium and dielectric behavior of the SF and sSF models.\cite{kohns_relative_2020,marx_phase_2022,marx_vapor-liquid_2023} To the best of our knowledge, the properties of the liquid--vapor interface for the sSF model have yet to be studied.

In this study, we investigate whether a small shift of the dipole from the molecular center is sufficient to drive interfacial polarization and generate a measurable electric field. To address this question, we perform MD simulations of the liquid--vapor interface using the sSF model. Our analysis uncovers the mechanisms of orientational polarization at the interface and quantifies the spontaneously generated local electric field across a wide range of dipole strengths and shifts. To interpret these results, we employ a simple image dipole framework that clarifies the origin of the angular distribution functions in the liquid and vapor regions adjacent to the interface. We start by presenting the simulation setup and methodology for studying the interface in the next section.

% Molecules can exhibit asymmetry within their structure through varying sizes of atoms and functional groups. Non-spherical, non-polar, neutral molecules show preferential orientation at interfaces in order to maximize their entropy.\cite{gubbins_molecular_1981,thompson_molecular_1979,thompson_structure_1981,eggebrecht_liquidvapor_1987-1}  Enders et al. hypothesized that non-spherical, polar molecules would exhibit a crossover from an entropy dominated orientational distribution to an energy driven one as the degree of asymmetry decreased and the dipole strength increased.\cite{enders_molecular_2004} While Johnson et al. looked at the bulk dielectric behavior of dipolar spheroids,\cite{johnson_dielectric_2015} nobody has yet looked at the interfacial behavior as far as we know.

\section{Methodology}
In this section, we outline the simulation setup and methodology for analyzing the properties of the liquid--vapor interface for the shifted Stockmayer fluid model.

\subsection{Molecular Dynamics}
\begin{figure*}[hbt!]
    \centering
    \includegraphics[width=1.0\linewidth]{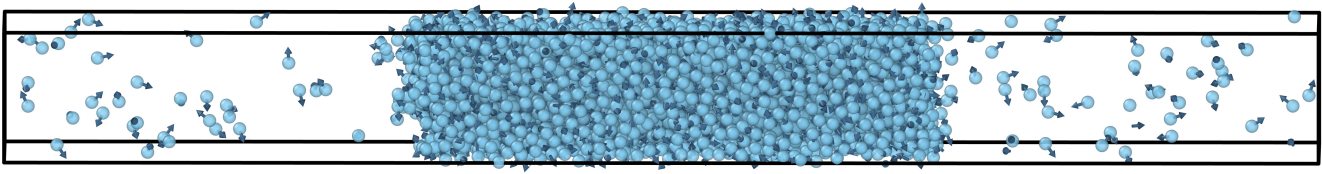}
    \vspace{-0.5cm}
    \caption{Snapshot of liquid--vapor equilibrium simulation using the sSF model. Blue spheres represent the Lennard--Jones particles and yellow arrows represent the point dipoles which are interior to the Lennard-Jones particles. Visualizations are made in OVITO.\cite{ovito}}
    \label{fig:md-setup}
\end{figure*}
To study the interfacial structure in asymmetric polar fluids, we utilize a modified version of the Stockmayer fluid (SF) model sometimes referred to as the \textit{shifted} Stockmayer fluid (sSF).\cite{langenbach_co-oriented_2017} In this model, molecules are represented by hard spherical particles that contain a point-dipole. In the SF, the point dipole is located at the center of the hard sphere, whereas in the sSF the dipole can be shifted off the center. The hard-sphere and van der Waals interactions are captured using the 12-6 Lennard-Jones (LJ) potential.
\begin{equation}
\label{eq:lj}
U_{\text{LJ}}(r_{ij}) = 
\begin{cases} 
4\epsilon \left[ \left( \frac{\sigma}{r_{ij}} \right)^{12} - \left( \frac{\sigma}{r_{ij}} \right)^6 \right]+S\,, & \text{if } r_{ij} \leq r_c \\
0\,, & \text{if } r_{ij} > r_c .
\end{cases}
\end{equation}
Here, $\epsilon$ is the interaction strength, $\sigma$ is the particle size, and $r_{ij}=||\*r_{ij}||$ is the distance between particles $i$ and $j$. We utilize the truncated and shifted LJ potential with a cutoff of $r_c=2.5\sigma$ to decrease the computational cost while still allowing for a liquid--vapor split. $S$ then provides the shift necessary such that $U_\mathrm{LJ}(r_c)=0$. We note that this choice is inconsequential since we only require the van der Waals interaction to naturally create the interface within our simulation and not to provide any quantitative description or comparison to previous simulations or real fluids. Also note that the original work by Langenbach and co-workers used the full LJ potential. As such, quantitative differences are to be expected. The dipole--dipole interaction potential is given by,
\begin{equation}\label{eq:dipole}
U_\text{dipole}(\pmb{\mu}_i,\pmb{\mu}_j,\*r_{ij}) = \frac{\pmb{\mu}_i\cdot\pmb{\mu}_j}{r_{ij}^3}-\frac{(\pmb{\mu}_i\cdot\mathbf{r}_{ij})(\pmb{\mu}_j\cdot\mathbf{r}_{ij})}{r_{ij}^5}\,,
\end{equation}
where $\pmb{\mu}_i$ is the dipole moment vector of particle $i$, and $\*r_{ij}$ is the displacement vector between dipoles $i$ and $j$.\cite{allen_computer_2017} To represent the sSF model, we separate the LJ and dipole potential between an LJ (real) and dipole (ghost) particle. The location of the dipole particle is offset by a distance $d$ from the center of the LJ particle. Thus, the location of the dipole particle can be written with respect to the LJ particle coordinate according to $\*r'_i=\*r_i+\hat{\pmb{\mu}}_id$. Note that this also requires that the dipole moment will always be pointing away from the center of the LJ particle, parallel to the bond vector between the two particles. As such, the LJ+dipole molecule must be rigid, with no internal translation or rotation. To enforce this constraint within our MD simulations, we utilize the rigid small-molecule NVT integrator that is freely available in LAMMPS.\cite{kamberaj_time_2005} To compute the long-ranged dipole--dipole interactions, we utilize the particle-particle particle-mesh (PPPM) solver.\cite{toukmaji_efficient_2000,cerda_p3m_2008} We adjust the real-space cutoff for each simulation independently to maximize the efficiency. The cutoff is typically between $8-10\sigma$.

To generate a natural interface, we employ a strategy commonly used for studying liquid--vapor equilibrium. We simulate 3000 molecules total (i.e., 3000 LJ particles and 3000 dipoles) in a box of dimensions $L_x,L_y,L_z=10\sigma,10\sigma,100\sigma$. The extended $z$-dimension allows for a liquid slab located in the center of the box that is approximately $40-50\sigma$ in length, with vapor on either side. This style of system setup has been widely used for studying properties of liquid--vapor and liquid-liquid interfaces with both classical and ab-initio MD simulations,\cite{jungwirth_specific_2006,matsumoto_study_1988,matsumoto_molecular_1989,cendagorta_surface_2015,yeh_structure_2001,martins-costa_electrostatics_2023,medders_dissecting_2016,moberg_temperature_2018,kuo_ab_2004,kathmann_understanding_2011,tobias_simulation_2013,kathmann_electronic_2008,mundy_hydroxide_2009,dodia_structure_2019,horvath_vapor-liquid_2013,mecke_molecular_1997,taylor_molecular_1996} and has also been used to study the liquid--vapor equilibrium of the regular Stockmayer fluid.\cite{moore_liquid-vapor_2015,samin_vapor-liquid_2013,eggebrecht_liquidvapor_1987,enders_molecular_2004,paul_liquidvapor_2003,mecke_molecular_2001} We provide an example snapshot from our slab simulations in Figure \ref{fig:md-setup}. We assign a mass of $m=1$ to the LJ particle and $m=0.001$ to the dipole particle. In addition, we turn off all LJ interactions involving the point dipole particles. Unless otherwise stated, we set $k_B=1$, $T=1$, $\sigma=1$, and $\epsilon=1$. Within a given simulation, all dipole particles have the same dipole moment magnitude, $\mu=\|\boldsymbol{\mu}_i\|$. We use the \texttt{rigid/nvt/small} integrator in LAMMPS with a timestep of $0.005\tau$ and a damping parameter of $0.5\tau$, where $\tau=\sqrt{m\sigma^2/\epsilon}$.\cite{kamberaj_time_2005}

We create a randomized initial configuration using Packmol.\cite{martinez_packmol_2009} Then, we create the liquid slab in the center of the simulation box by running a short simulation with a dragging force applied in the $z$-direction that is antisymmetric about $z=L_z/2$. Once the liquid slab is stabilized, we turn off the dragging force and allow the liquid--vapor equilibrium to develop for $5\times 10^4$ timesteps. Following the equilibration run, we perform a production run for $4\times 10^6$ timesteps, where we output the particle coordinates and dipole orientations every 100 timesteps for analysis.

Our main focus of this work is to analyze the interfacial properties of the sSF. Thus, it behooves us to utilize a rigorous definition of the interface, which can undulate and drift throughout the course of our simulations. By computing the location of the interface at each simulation frame, we can shift the coordinates during analysis to maintain a consistent frame for computing spatially varying profiles. We define the location of the interface according to the \textit{instantaneous interface}, as defined by Willard and Chandler.\cite{willard_instantaneous_2010,tarazona_intrinsic_2012} Specifically, we center Gaussians of form $f(r)=(2\pi\eta^2)^{-3/2}e^{-r^2/2\eta^2}$ at the location of each \textit{LJ} particle. We sum all of the Gaussians, accounting for periodic boundary conditions, to generate a continuous three-dimensional density field. We then define the \textit{left} and \textit{right} interfaces by computing the two density iso-surfaces $z_l(x,y)$ and $z_r(x,y)$, respectively. The target density is chosen to be exactly halfway between that of the liquid and the vapor, as determined from the computed density field. To obtain a scalar definition of the location of the two interfaces, we compute their averages via integration using $\bar{z}_l=(L_xL_y)^{-1}\int dx\int dy \,\,z_l(x,y)$ and $\bar{z}_r=(L_xL_y)^{-1}\int dx\int dy \,\,z_r(x,y)$. Visualizations of each step of this workflow are provided in the ESI$^\dagger$.

Finally, during analysis we use the instantaneous interface position to shift the particle coordinates. We believe this method is more robust than zeroing the system center of mass, as  the presence of more vapor particles on one side of the liquid than the other can significantly affect the interface location. All trajectory analysis was performed using custom analysis modules implemented in the open-source MDCraft software.\cite{ye_mdcraft_2024}

\section{Results and discussion}

Here, we provide a detailed analysis of the density and electrostatic profiles at the liquid--vapor interface for a variety of conditions. In general, we span from weak to strong dipole moments, $\mu\in1.0-2.0$, and small to large offsets, $d\in0.0-0.25$, and study the effects on the liquid density and interfacial polarization. For some additional context, we mention briefly here how water maps onto the parameters of this model. Namely, choosing $\sigma=2.75$\,\AA, $\mu_0=1.8$\,D, and $T=300$\,K yields a reduced dipole moment of $1.94$. Modeling the water very roughly as a spherical molecule, we can compute the effective dipole shift as the distance between the center of geometry and the center of charge as approximately 0.10\,\AA  (or 0.036$\sigma$). Other molecules, such as refrigerants, typically have much large dipole shifts, reaching up to 0.2$\sigma$.

We rely on angular distribution functions to provide a detailed molecular picture and explanation of the observed interfacial polarization. 

\subsection{Density Profiles}

\begin{figure}[h!]
    \centering
    \includegraphics[width=1\linewidth]{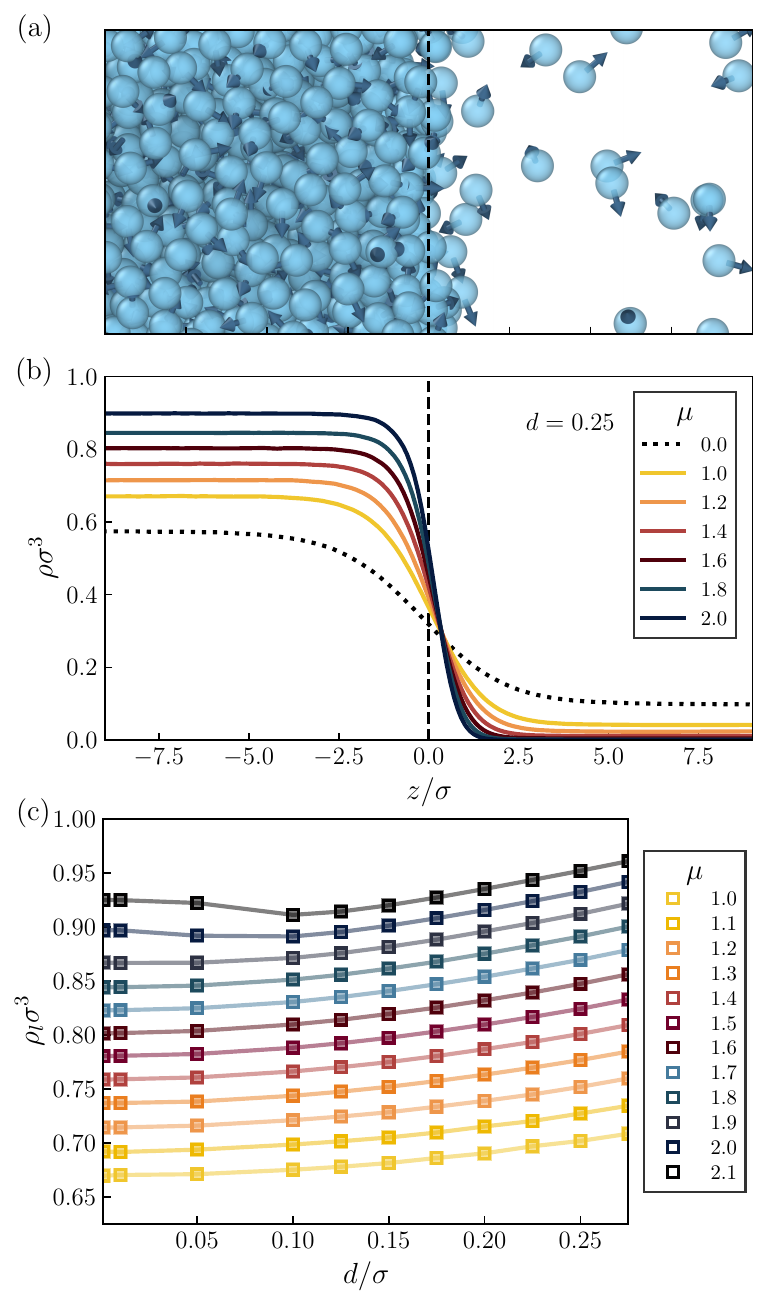}
    \vspace{-0.5cm}
    \caption{Interfacial density profiles for various $\mu$ and $d$. The interface is located between a bulk region of liquid on the left and a bulk region of vapor on the right. (a) Simulation snapshot with the average of the instantaneous interface plotted as a dashed line. (b) Density profiles with $d=0.25\sigma$ and $T=1$ for various values of $\mu$, with the dotted line representing the reference LJ fluid with no dipoles. (c) Coexistence liquid densities at $T=1$ for various values of $\mu$ and $d$.}
    \label{fig:density}
\end{figure}

We first compute the density profiles across the liquid--vapor interface using Equation \eqref{eq:density} given as,
\begin{equation}\label{eq:density}
    \rho(z)=\frac{1}{L_x L_y}\sum_{i=1}^N \delta(z-z_i)\,,
\end{equation}
where $\delta(x)$ is the Dirac delta function. In practice, the coordinates are binned into narrow windows instead of using an exact delta function. Density profiles at various dipole strengths are provided in Figure \ref{fig:density}. As expected, Figures \ref{fig:density}b and c show that the density increases and the interface becomes sharper as $\mu$ increases. Note that this effect is not unique to shifted dipoles and occurs for any value of $d$. 

 The effects of $\mu$ and $d$ on the coexistence liquid density are further summarized in Figure \ref{fig:density}c. Here, we see that the liquid density increases as $d$ increases for a given value of $\mu$. We attribute the increase in density to a strengthening of dipolar interactions, as the introduction of a shift deepens the potential energy minimum in the sSF potential, thus allowing the particles to pack more closely. When $\mu>1.8$, increasing $d$ first results in a decrease in density, followed by an increase in density that follows the same trend as weaker dipole moments. This non-monotonic behavior is caused by the presence of a ferroelectric transition, which has been reported for dipolar hard spheres at high density.\cite{groh_ferroelectric_1994,bartke_dielectric_2006,johnson_dielectric_2015,pounds_are_2007,weis_ferroelectric_1993} When $d=0$, the dipoles prefer to align head to tail to minimize their energy. When $\mu$ is low, this preference is not strong enough to overcome thermal fluctuations and cause any spontaneous orientation in the liquid. However, when the dipole moment becomes strong enough, the dipoles spontaneously align, much like the spontaneous magnetization observed in ferromagnetic materials. This ferroelectric transition also causes the liquid to contract, resulting in an increased liquid density. As $d$ increases, the perfect head-to-tail configuration of dipoles becomes less favorable relative to a slightly rotated configuration, and the ferroelectric transition is disrupted. A similar effect was also observed by Johnson et al. in their study of dipolar spheroids, where the ferroelectric transition was suppressed for both oblate and prolate spheroidal particles.\cite{johnson_dielectric_2015}
 
 Since the shift places the dipoles closer to the particle surface, the optimal configuration for two dipoles is different than the regular Stockmayer potential. The dipole--dipole interaction energy is determined by both the distance and relative angle of the two dipoles. At large dipole shifts, the energy can become lower than the perfect head-to-tail configuration by rotating one of the particles slightly. This puts the dipoles closer together while sacrificing a small amount of attraction due to the non-optimal angle. The net effect is a lower overall energy due to the dipole interaction. Thus, at small but finite $d$, the ferroelectric transition is diminished but not fully suppressed, resulting in a decreasing density. At some value of $d$, the ferroelectric transition is suppressed completely, and further increasing $d$ results in an increased density for the same reasons described in the low $\mu$ case. A more detailed description of the liquid structure and angular distribution functions is provided in our companion study on the bulk equilibrium properties of the sSF model (Venkatachalam et al. 2025).

% \begin{figure}[h!]
%     \centering
%     \includegraphics[width=1\linewidth]{figures-pdf/rho_vs_d.pdf}
%     \vspace{-0.5cm}
%     \caption{Coexistence density versus offset $d$ for various values of $\mu$. Lines are only provided to guide the eye.}
%     \label{fig:rho_vs_d}
% \end{figure}

\begin{figure}[b!]
    \centering
    \includegraphics[width=1\linewidth]{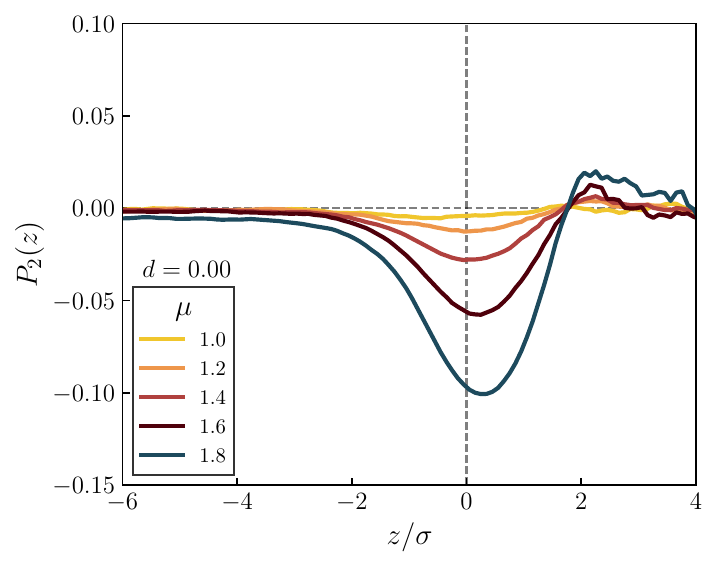}
    \vspace{-0.5cm}
    \caption{Nematic order profiles with interface surface normal as reference director. Profiles are plotted for no shift ($d=0$) for various values of dipole moment $\mu$. Note that $P_2(z)<0$ indicates alignment parallel to the interface (perpendicular to the normal vector). The dashed lines are provided to guide the eye.}
    \label{fig:S_z_mu}
\end{figure}

\subsection{Nematic and Polar Order}

\begin{figure*}[t!]
    \centering
    \includegraphics[width=1\linewidth]{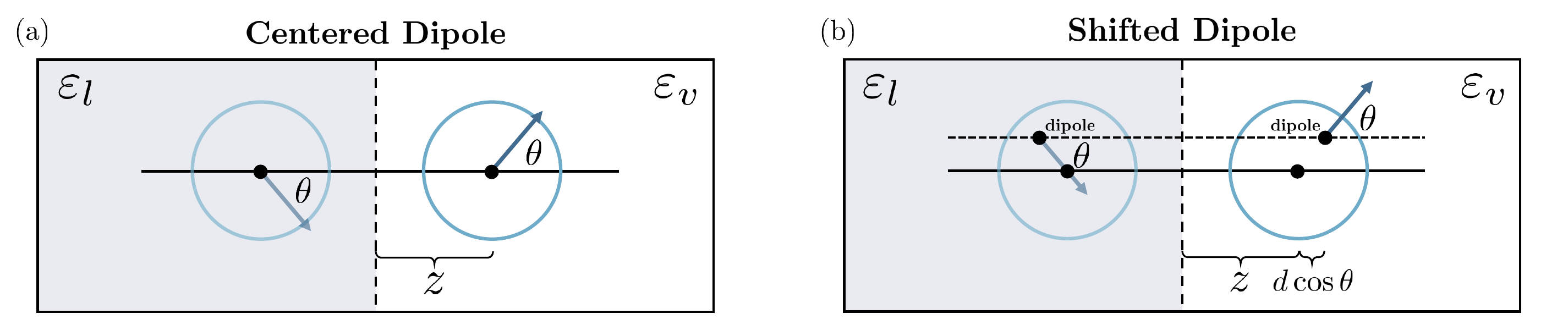}
    \vspace{-0.7cm}
    \caption{Image dipole construction at a liquid--vapor interface of the (a) Stockmayer fluid and (b) shifted Stockmayer fluid.}
    \label{fig:image-dipole}
\end{figure*}
 
A regular Stockmayer fluid with no dipole shift will necessarily have zero polar order throughout the fluid due to symmetry (assuming no ferroelectric transition occurs). The lack of polar order means that there can be no interfacial polarization and therefore no electric field or potential difference between the two phases. Despite the lack of polar order, we can still make an interesting observation about the nematic order at the interface for the regular SF model. Specifically, we take the unit normal of the interface to be a reference director, and we compute the relevant component of the nematic order parameter via the 2nd Legendre polynomial of orientation as
\begin{equation}
    P_2(z)=\frac{1}{2}\left[3\left\langle\cos^2\theta\right\rangle(z)-1\right]\,,
\end{equation}
where $\theta$ is the angle between a given dipole and the normal vector of the interface. The average $\langle\cdots\rangle(z)$ is computed within a narrow window around $z$. In words, $P_2(z)$ is a measure of how aligned the dipoles are with the surface normal as a function of distance from the interface. There are 3 interesting cases: (1) $P_2=-1/2$ if all dipoles are perpendicular to the surface normal, (2) $P_2=1$ if all dipoles are parallel or antiparallel to the surface normal, (3) $P_2=0$ if the dipoles are randomly oriented with respect to the surface normal. In subsequent discussion, we refer to $P_2$ as the nematic order for simplicity, but it should be understood that it is only a component of the nematic order with an assumed nematic director parallel to the surface normal. This quantity has been used in several previous theoretical and simulation studies of Stockmayer fluids at interfaces.\cite{moore_liquid-vapor_2015,mecke_molecular_2001,eggebrecht_liquidvapor_1987,eggebrecht_liquidvapor_1987-1,paul_liquidvapor_2003,teixeira_density-functional_1991,teixeira_orientation_2002}

\begin{figure}[b!]
    \centering
    \includegraphics[width=1\linewidth]{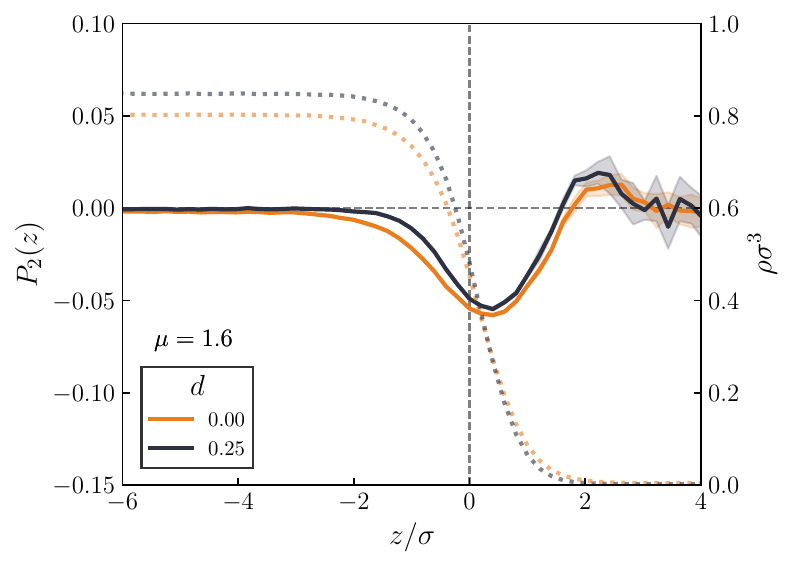}
    \vspace{-0.5cm}
    \caption{Nematic order profile (solid lines) from extended simulations for $\mu=1.6$ and $d=0,\,0.25$. The shaded areas represent 95\% confidence intervals computed from block averaging, and the dotted lines are the corresponding density profiles.}
    \label{fig:S_z_mu_long}
\end{figure}

We plot the nematic order profile for the case of no dipole shift in Figure \ref{fig:S_z_mu}. It is clear from $P_2(z)<0$ that there is a preference for the dipoles to lie in the interfacial plane, even with perfectly symmetric spherical molecules. This can be understood by considering image dipole interactions generated at the interface.\cite{teixeira_orientation_2002,frodl_thermal_1993,ruiz-lopez_molecular_2020,martins-costa_solvation_2015} Assuming that the interface can be approximated by a discontinuous jump between two homogeneous phases with disparate dielectric constants, molecules near the interface will experience interactions with an image dipole on the opposite side that has the same polar angle but is rotated azimuthally by $\pi$. This scenario is depicted in Figure \ref{fig:image-dipole}a. The resulting interaction energy for a particle near the interface is 
\begin{equation}\label{eq:image-dipole}
    U(z,\theta) = \frac{\mu^2}{16}\frac{\epsilon_\alpha-\epsilon_\beta}{\epsilon_\alpha(\epsilon_\alpha+\epsilon_\beta)}\frac{1+\cos^2\theta}{z^3}\,,
\end{equation}
where $\epsilon_\alpha$ is the static dielectric constant of the host phase, $\epsilon_\beta$ is the dielectric constant of the coexisting phase, $\theta$ is the polar angle relative to the surface normal, and $z$ is the distance from the interface. Note that the net sign of the interaction depends on whether the dipole is within the high dielectric or low dielectric medium. We can use Equation \eqref{eq:image-dipole} to understand the orientational behavior of molecules on both the liquid and vapor sides of the interface. As we will later show, this analysis can also be used to predict the shape of the angular probability distributions, even in the presence of a dipole shift.

\begin{figure*}[hbt!]
    \centering
    \includegraphics[width=1\linewidth]{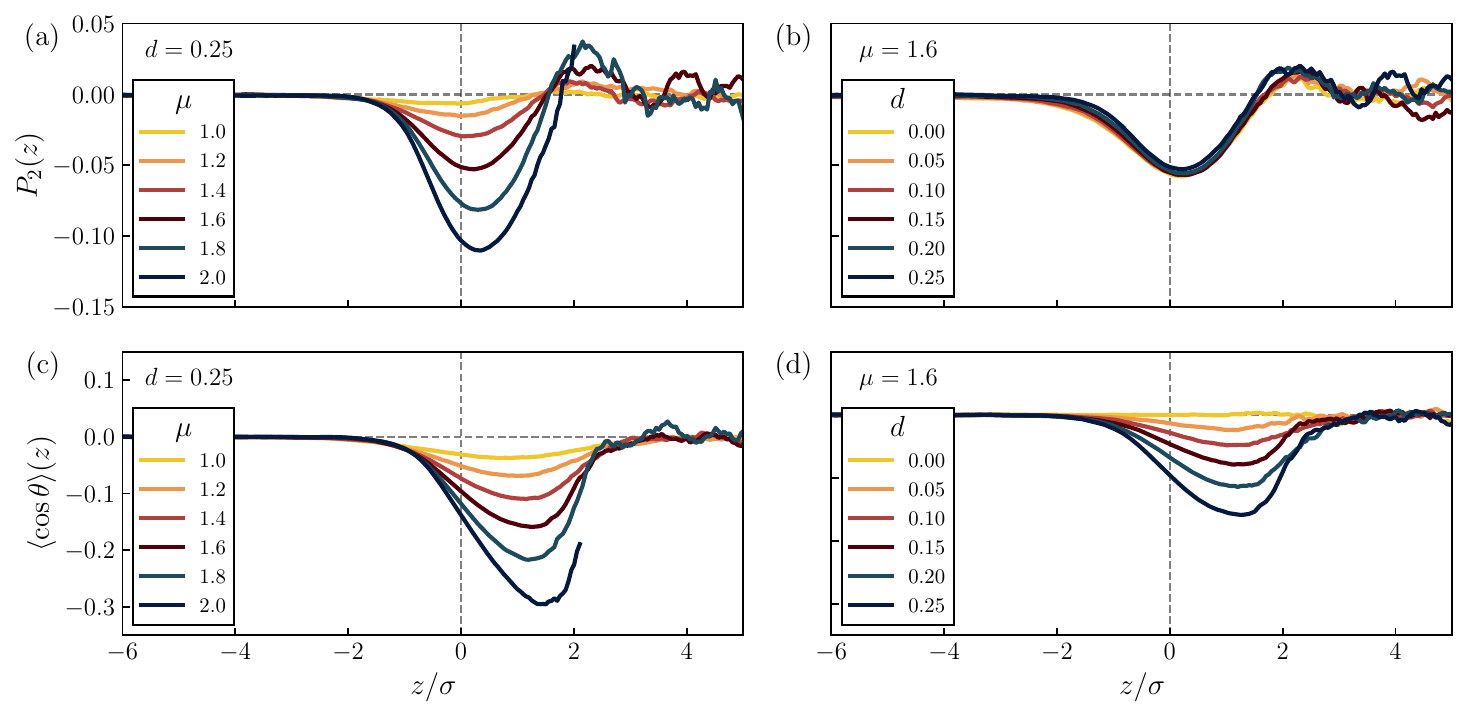}
    \vspace{-0.5cm}
    \caption{Spatially varying nematic ((a) and (b)) and polar ((c) and (d)) order parameters with respect to surface normal vector. (a)+(c) Profiles for a shift of $d=0.25$ for various values of dipole moment $\mu$. (b)+(d) Profiles for a dipole moment of $\mu=1.6$ for various values of shift $d$. We omit averages computed from less than 100 data points due to significant statistical uncertainty. Note that $m(z)<0$ and $m(z)>0$ indicate dipoles pointing towards the liquid and vapor, respectively. The dashed lines are provided to guide the eye.}
    \label{fig:m_z_mu_d}
\end{figure*}

For a molecule in the liquid ($\epsilon_\alpha>\epsilon_\beta$), the energy in Equation \eqref{eq:image-dipole} is minimized when $\theta=\pi/2$, indicating that the molecules will, on average, prefer to lie parallel to the interface. Alternatively, for a molecule in the vapor ($\epsilon_\alpha < \epsilon_\beta$), the energy is minimized when $\theta=0$ and $\pi$, and the molecule will prefer to point perpendicular to the interface. Several studies using classical density functional theory (cDFT) and liquid state theory (LST) have predicted an S-shaped $P_2(z)$ profile across the liquid--vapor interface, where molecules on the liquid side are parallel to the interface and molecules on the vapor side are perpendicular.\cite{eggebrecht_liquidvapor_1987-1,frodl_bulk_1992,frodl_thermal_1993,iatsevitch_structure_2000} On the other hand, several molecular dynamics simulation studies predicted only a parallel orientation ($P_2<0$) on both sides of the Gibbs dividing surface (GDS).\cite{moore_liquid-vapor_2015,mecke_molecular_2001,enders_molecular_2004,eggebrecht_liquidvapor_1987} To the best of our knowledge, this disagreement between theory and simulation has still not been resolved. The two main hypotheses for the disagreement are (1) that the theoretical frameworks are unable to fully describe the essential physics of the density and dipole correlations, and (2) that the profiles computed from simulation carry too much statistical uncertainty, particularly in the dilute vapor phase.

The simulation results in Figure \ref{fig:S_z_mu} agree with previous MD simulations of Stockmayer fluids in that $P_2(z)$ is negative across the interface, with the minimum located near the GDS. In contrast to previous works, our simulations seem to produce a pronounced peak within the vapor region when the dipole moment is high ($\mu\ge1.6$). However, significant noise once again makes the interpretation difficult. To verify this behavior with statistical certainty, we conducted simulations with $\mu=1.6$ and $d=0,\,0.25$ for $10^7$ timesteps. We saved every 100th frame for analysis, resulting in a total of $10^5$ frames. Treating the two interfaces as independent, we were able to double the amount of data and obtained the profile of the nematic order using a total of $2\times 10^5$ frames. We broke the data into 10 blocks of $2\times10^4$ frames each and used the standard error of the mean to estimate the uncertainty in the profiles. The resulting nematic order parameter profiles with 95\% confidence intervals are plotted in Figure \ref{fig:S_z_mu_long}. These show a statistically significant peak within the vapor phase, which agrees qualitatively with the predictions of cDFT and LST, as well as the image dipole result based on Equation \eqref{eq:image-dipole}. However, we note that cDFT and LST predict that the transition from negative to positive $P_2(z)$ occurs at the GDS, whereas our calculations show that $P_2(z)$ does not become positive until around $2\sigma$ outside the GDS.

Introducing a shift to the dipole has little effect on the spatially varying nematic order parameter, which we plot in Figure \ref{fig:m_z_mu_d}b. We can modify Equation \eqref{eq:image-dipole} to include the dipole shift, which results in
\begin{equation}\label{eq:image-dipole-shifted}
    U(z,\theta) = \frac{\mu^2}{16}\frac{\epsilon_\alpha-\epsilon_\beta}{\epsilon_\alpha(\epsilon_\alpha+\epsilon_\beta)}\frac{1+\cos^2\theta}{\left(z+d\cos\theta\right)^3}\,,
\end{equation}
where $d$ is the dipole shift. The shift effectively couples the dipole angle with the distance of the dipole from the interface. This scenario is depicted in Figure \ref{fig:image-dipole}b. Dipoles pointing towards the interface will have stronger interactions with their images than those that are pointing away. Note that the effect of the shift diminishes as $z$ grows larger than $d\cos\theta$ and the energy approaches that of Equation \eqref{eq:image-dipole}. The asymmetry introduced by the shift leads to a preferential orientation of the dipoles at the interface, which is most easily seen through the polar order profile and the shape of the angular distribution functions in the interfacial region.

\begin{figure*}[hbt!]
    \centering
    \includegraphics[width=1\linewidth]{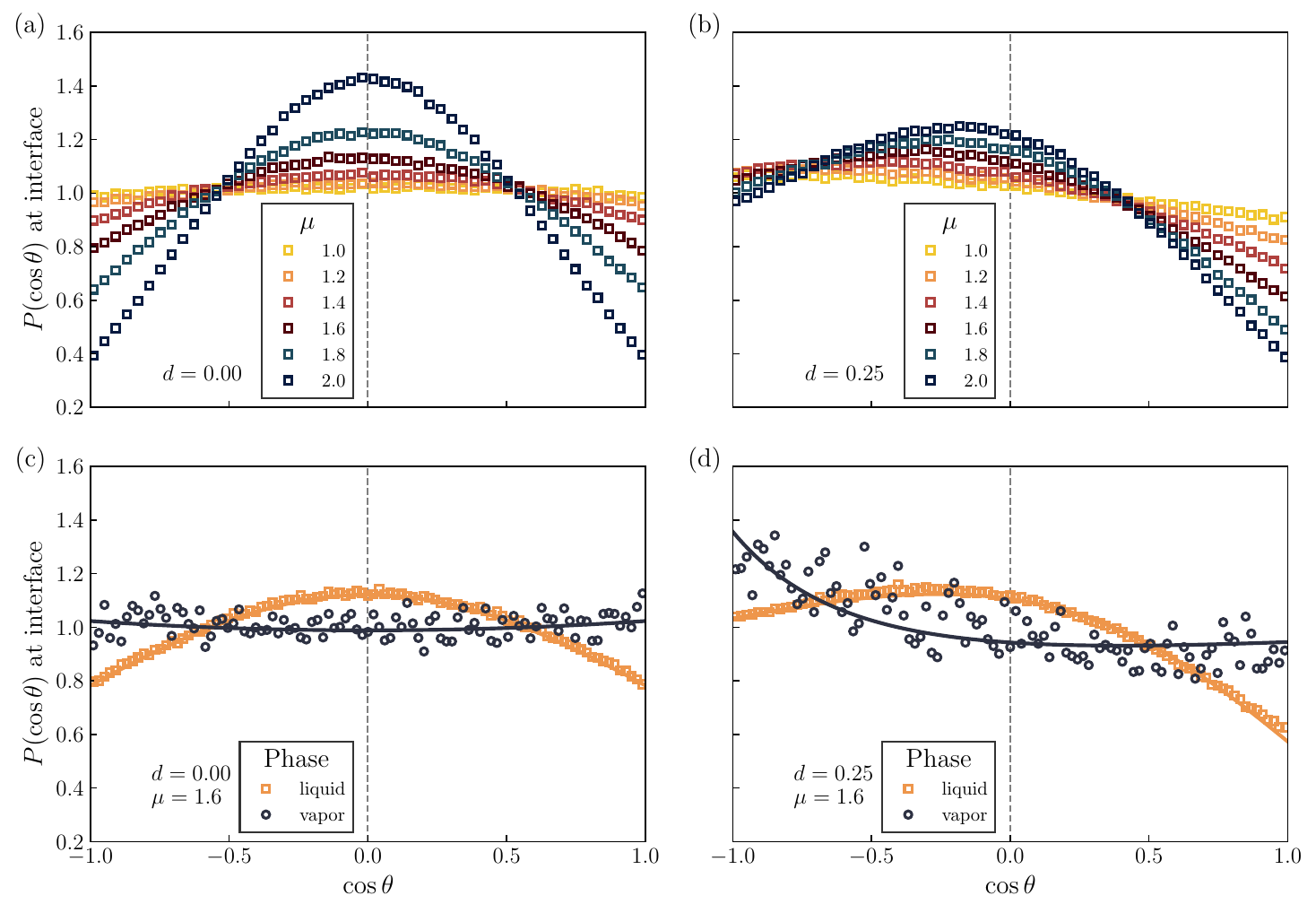}
    \vspace{-0.8cm}
    \caption{Angular distribution functions in the interfacial region for various dipole moments $\mu$ and dipole offsets $d$. Distributions in (a) and (b) are extracted from the entire interfacial region within $-0.5<z<2.0$. Distributions in (c) and (d) are broken into separate regions for the liquid and vapor phases. Symbols represent MD data and curves represent fits to Equations \eqref{eq:image-dipole} and \eqref{eq:image-dipole-shifted} for (c) and (d) respectively. The distributions are normalized and then multiplied by a constant factor of $2$. Recall that $\cos\theta<0$ and $\cos\theta>0$ correspond to molecules pointing towards the liquid and vapor, respectively.}
    \label{fig:P_cos_theta}
\end{figure*}

Similar to $P_2(z)$, we define the spatially varying polar order parameter $m(z)$ in Equation \eqref{eq:polar}. We use the coordinates of the LJ particles to determine the positions and the dipole vectors for the orientations.
\begin{equation}\label{eq:polar}
    m(z)=\langle\cos\theta\rangle(z)=\langle \hat\mu_z\rangle(z)\,.
\end{equation}
Again, $\theta$ is the angle between the dipole and the surface normal which is the unit vector in the $z$-direction, and theaverage $\langle\dots\rangle(z)$ is computed within a narrow window around $z$. In words, $m(z)$ is a measure of the average orientation of the molecules, specifically in the $z$-direction. A value of $m(z)=0$ means that the molecules have no \textit{net} preference to point towards the liquid or vapor, while $m(z)<0$ and $m(z)>0$ indicate preferences to orient towards the liquid and vapor regions, respectively. Note that $m(z)=0$ with $P_2(z)\ne0$ is achieved by having a non-uniform distribution on $\cos\theta$ that is still symmetric about $\theta=\pi/2$, which is exactly the case for the regular Stockmayer fluid.

We plot the spatially varying polar order for a dipole shift of $d=0.25$ in Figure \ref{fig:m_z_mu_d}c. A shift in the dipole induces an orientational order normal to the interface across a wide range of dipole moments. The strength of the ordering increases with dipole moment as the electrostatic interactions become stronger and a denser solvation environment becomes more favorable. In Figure \ref{fig:m_z_mu_d}d, we show that the orientational order also becomes more pronounced as the dipole shift increases for a fixed dipole moment. These observations are also consistent with the image dipole construction of Equation \eqref{eq:image-dipole-shifted}. Namely, a finite $d$ skews the minimum of the energy due to image dipole interactions for molecules in both the liquid and vapor phases. In both cases, the molecules will show a preference to point towards the liquid phase when $d>0$, resulting in a purely negative $m(z)$ across the interface.

Together, Figures \ref{fig:m_z_mu_d}a and \ref{fig:m_z_mu_d}c paint an almost complete picture of the molecular orientations across the liquid--vapor interface for the sSF model. The liquid features $P_2(z)<0$ and $m(z)<0$, indicating that molecules are largely oriented in the plane of the interface, with a slight preference to point towards the liquid. On the other hand, the vapor features $P_2(z)>0$ and $m(z)<0$, indicating that molecules are largely oriented perpendicular to the interface, also with a preference to point towards the liquid.

Providing further quantitative insights, we plot the probability distribution of $\cos\theta$ within the interfacial region in Figure \ref{fig:P_cos_theta}. We denote these as \textit{angular distribution functions}. The interfacial region is chosen as $-0.5<z<2.0$ based on the profiles in Figure \ref{fig:m_z_mu_d}. The distributions in Figure \ref{fig:P_cos_theta}a show that, for the SF, the preference to lie in the interfacial plane becomes stronger with increasing dipole moment, indicated by the sharpening of the peak around $\cos\theta=0$. The picture becomes more complicated when the dipole is shifted off-center, as shown for $d=0.25$ in Figure \ref{fig:P_cos_theta}b. For weak dipole moments, the distribution becomes skewed towards $-1$. As the dipole moment increases, the skew increases, and a prominent peak appears within $-1<\cos\theta<0$. The shape of this distribution explains how molecules with high $\mu$ and high $d$ have significantly negative $P_2(z)$ and $m(z)$ near the interface. The negative skew in the distributions supports the conclusion that the molecules prefer to point towards the liquid. Due to the way we constructed our molecules, this also implies that the molecules are positioning their point-dipoles closer to the liquid phase than the vapor phase. This orientation allows for the dipoles to interact more strongly with dipoles just below the surface layer, providing some additional solvation from liquid-phase particles.

As mentioned previously, Equations \eqref{eq:image-dipole} and \eqref{eq:image-dipole-shifted} can be used to make predictions about the angular distributions on the liquid and vapor sides of the interface. Figures \ref{fig:P_cos_theta}c and \ref{fig:P_cos_theta}d show remarkable qualitative agreement between the MD simulations and the theoretical predictions. The lines were computed through a fitting procedure where the parameters of the distribution (i.e., $\varepsilon$, $\mu$, $d$, $z$, etc.) were varied to minimize residuals. The open symbols were computed by splitting the interfacial region into a \textit{liquid side} and a \textit{vapor side}, and accumulating dipole orientations in the two separate regions. While the liquid--vapor interface itself is diffuse, we choose to define the transition from liquid to vapor at $z\approx 2\sigma$ based on the location where $P_2(z)$ crosses from negative to positive in Figures \ref{fig:S_z_mu}, \ref{fig:S_z_mu_long}, and \ref{fig:m_z_mu_d}. We do this to remain in line with the predictions of the image-dipole method which predicts that $P_2$ is negative in the liquid and positive in the vapor. We don't expect quantitative agreement as we have a diffuse interface and significant dipole correlation effects that are completely ignored by treating dipoles as independent particles immersed in a homogeneous medium with a dielectric discontinuity. Amazingly, the qualitative behavior can still be captured for both the liquid and vapor, whether or not there is a dipole offset.

\subsection{Interfacial Electric Field and Potential}

The primary objective of this work is to determine if molecular asymmetry in polar fluids is sufficient to induce a significant interfacial electric field. To this end, we discuss in the following section the electric properties of the interface including the electric field and the electrostatic potential profile. To compute the spatially varying electric field, we start from the differential form of Gauss's Law,
\begin{equation}\label{eq:gauss}
    \varepsilon_0 \nabla\cdot \*E = -\nabla \cdot \*P +\rho_f\,,
\end{equation}
where $\varepsilon_0$ is the vacuum permittivity, $\*E$ is the electric field, $\*P$ is the polarization density, and $\rho_f$ is the \textit{free} charge. In our case, no free ions are present such that $\rho_f=0$. In addition, the system is infinitely vast in the dimensions parallel to the interface, implying that $\*E_{||}$, $\*P_{||}$, and their derivatives are identically 0 in the absence of a bulk ferroelectric transition. Thus, Equation \eqref{eq:gauss} reduces to the simple relation $\varepsilon_0 E_z(z)=-P_z(z)$, where $E_z$ and $P_z$ are scalar fields for the $z$-component of each quantity, which both vanish deep within the homogeneous vapor and liquid phases. This relation implies that the electric field points in the opposite direction of the polarization. Importantly, this provides a straightforward way to compute $E_z$ from MD simulation, since $P_z$ is easily computed with Equation \eqref{eq:polarization} given below,
\begin{equation}\label{eq:polarization}
    P_z(z) = \frac{1}{L_z L_y}\sum_{i=1}^N \delta(z-z_i)\mu_z^i\,,
\end{equation}
where $\mu_z^i$ is the $z$-component of the dipole moment of molecule $i$, and $z_i$ refers to the position of the dipole particle (not the LJ particle). In practice, dipoles are binned within a small region around $z$ and accumulated over the course of a long simulation run. The electrostatic potential profile, $\psi(z)$, is then computed from the electric field or the polarization by recognizing that $E_z(z)=-\partial_z \psi(z)$, which yields the following integral equation
% \begin{equation}\label{eq:potential}
%     \psi(z)-\psi(0)=-\int_0^z E_z(t)dt = \frac{1}{\varepsilon_0}\int_0^z P_z(t)dt\,,
% \end{equation}
\begin{equation}\label{eq:potential}
    \psi(z)-\psi_0=-\int_{z_v}^z E_z(t)dt = \frac{1}{\varepsilon_0}\int_{z_v}^z P_z(t)dt\,,
\end{equation}
% where $\psi(0)$ is the reference potential. In all plots, the vapor phase is taken as the reference with $\psi(0)=0$.
where $\psi_0$ is the reference potential, taken to be deep within the vapor region.

\begin{figure*}[hbt!]
    \centering
    \includegraphics[width=1.0\linewidth]{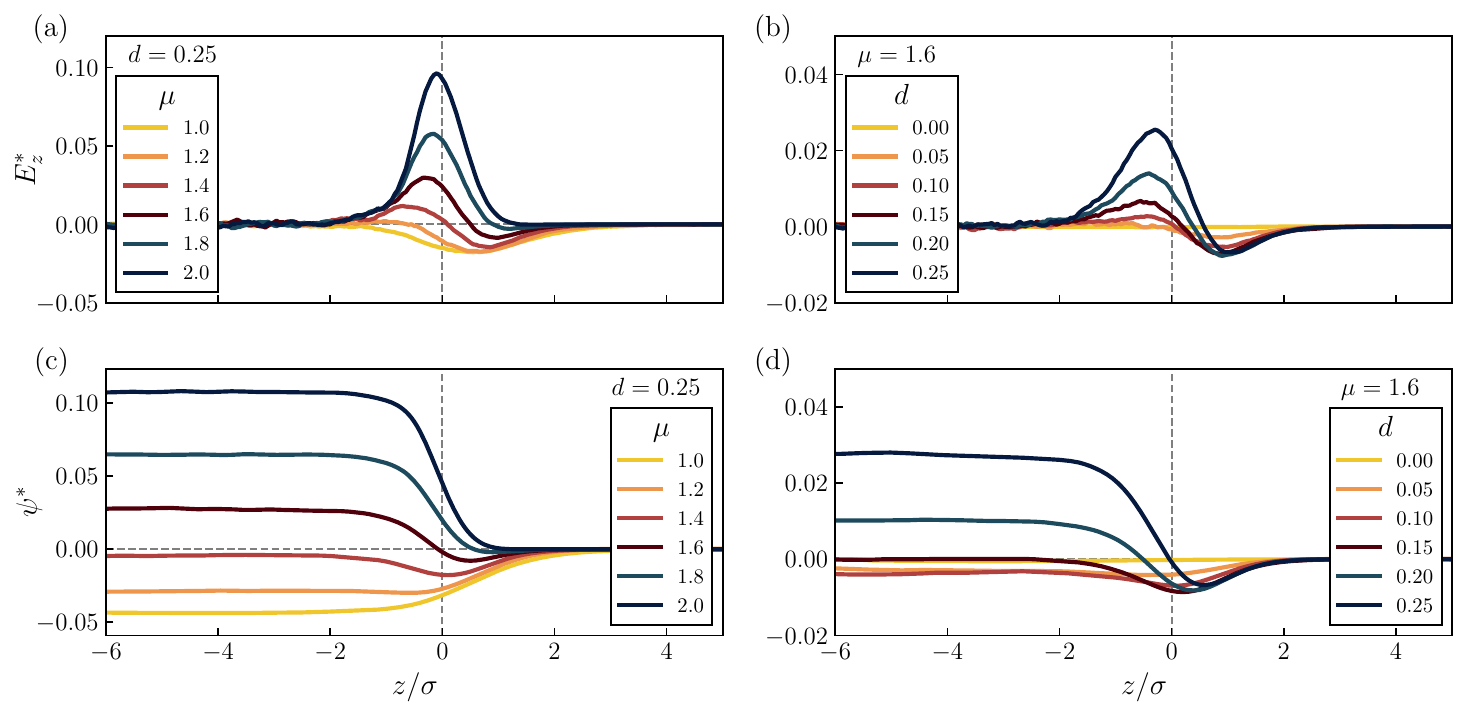}
    \vspace{-0.8cm}
    \caption{Spatially varying electric field ((a) and (b)) and electrostatic potential ((c) and (d)). (a)+(c) Profiles for a shift of $d=0.25$ for various values of dipole moment $\mu$. (b)+(d) Profiles for a dipole moment of $\mu=1.6$ for various values of shift $d$. The dashed lines are provided to guide the eye. The liquid phase is located at $z<0$ and the vapor phase at $z>0$.}
    \label{fig:E_psi_z}
\end{figure*}

We plot the reduced electric field, $E_z^*$, and the corresponding reduced electrostatic potential, $\psi^*$, in Figure \ref{fig:E_psi_z}. As expected, the electric field vanishes in the bulk phases and is finite at the interface. This interfacial field arises solely from the asymmetry of the polar molecules and the resulting polar order. Unexpectedly, the interfacial electric field $E_z^*$ can change sign. With increasing dipole moment strength, the electric field switches from negative to positive. This trend is also reflected in the electrostatic potential shown in Figures \ref{fig:E_psi_z}c where, for weak dipole moments, $\psi^*$ increases across the interface. For strong dipole moments, $\psi^*$ decreases. Figure Although we do not yet have a definitive physical explanation for this sign inversion, its existence is plausible given the complex interplay among liquid and vapor densities, interfacial width, dipole strength, and dipole shift. Here we provide one possible logical interpretation of this anomalous behavior.

The distribution functions in Figure \ref{fig:P_cos_theta}b show that increasing the dipole strength at a fixed, large shift enhances the tendency of interfacial molecules to orient up the density gradient toward the liquid phase ($-0.5<\cos\theta<0$). At sufficiently strong dipole moments, this preference becomes strong enough to produce a negative polarization and, correspondingly, a positive electric field (blue curves in Figure \ref{fig:E_psi_z}a and \ref{fig:E_psi_z}c). Thus, molecules with both a significant dipole shift and a strong dipole moment tend to generate polarization aligned with the density gradient.

By contrast, molecules with weak dipole moments exhibit a relatively flat angular distribution near the interface, which has two consequences. First, the weak preference for pointing toward the liquid is insufficient to generate appreciable negative polarization (positive electric field) at the weakest dipole moments studied. Molecules oriented toward the liquid point up the density gradient, placing their dipoles in a dense region where the polarization is averaged out by the random orientations of surrounding molecules. Molecules oriented toward the vapor point down the gradient, positioning their dipoles in a low-density region with few neighboring molecules. The combined effect is a net polarization of opposite sign to the density gradient (yellow/orange curves in Figures \ref{fig:E_psi_z}a and \ref{fig:E_psi_z}c).

The behavior of the electric field and electrostatic potential for various $d$ at fixed $\mu=1.6$ are given in Figures \ref{fig:E_psi_z}b and Figure \ref{fig:E_psi_z}d, respectively. The curves for $d=0$ confirm that the regular Stockmayer fluid produces no electric field at the interface and therefore no potential difference between the two phases. In general, the dipole strength sets the shape of the electric field at the the interface, while the dipole shift sets the magnitude. For example, at fixed $d=0.25$, the electric field goes from purely negative, to S-shaped, to purely positive as $\mu$ increases from 1.0 to 2.0. On the other hand, at fixed $\mu=1.6$, the shape remains S-shaped as $d$ increases while the positive peak on the liquid side of the interface increases in height. We provide further examples of this general observation for different regimes of $d$ and $\mu$ in the ESI$^\dagger$.

To quantify the strength of the interfacial electric field, we can substitute the properties of a real fluid. For water we have $\sigma=2.75$\,\AA, $\mu_0=1.8$\,D, and $\epsilon=k_B T$, which yields a reduced dipole moment of $\mu=1.94$. Thus, the simulations at $\mu=2.0$ can be reasonably interpreted within the context of the air--water interface. For $\mu=2.0$ and $d=0.25$, the reduced electric field at the interface is $E_z^*\approx0.10$ (Figure \ref{fig:E_psi_z}a), which corresponds to $E_z\approx16$\;MV/cm for the water properties listed above. This electric field is a reasonable order of magnitude for an air--water interface and agrees well with previous experiments,\cite{xiong_strong_2020} and molecular dynamics simulation using the ReaxFF/C-GeM forcefield.\cite{hao_can_2022} Any specific quantitative agreement should be interpreted cautiously since the sSF model is a simplified picture that ignores effects from electronic polarization, mean inner potential, hydrogen bonding, and steric effects of nonspherical molecules. We make this comparison simply to show that the magnitude of the interfacial electric field can reach experimentally relevant values.

\begin{figure}[hbt!]
    \centering
    \includegraphics[width=1.0\linewidth]{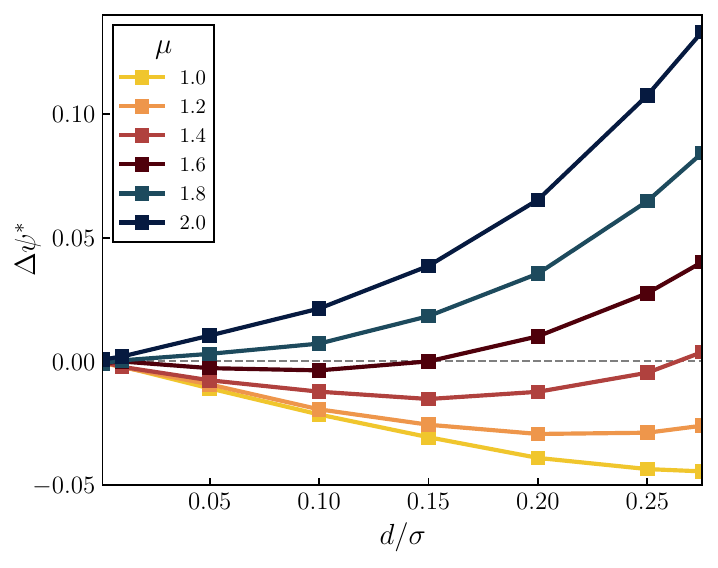}
    \vspace{-0.8cm}
    \caption{The potential difference between the bulk regions of vapor and liquid for various $\mu$ and $d$. $d=0$ corresponds to the regular Stockmayer fluid, which has no polarization and therefore no potential difference. $\Delta \psi^*=\psi^*_l-\psi^*_v$.}
    \label{fig:Delta-psi}
\end{figure}

The potential difference between the two phases across all combinations of $\mu$ and $d$ is given in Figure \ref{fig:Delta-psi}. The sSF model exhibits highly nontrivial behavior in the potential difference across the range of parameters studied here. For weak dipole moments, the dipole shift leads to a potential drop across the interface when going from the liquid to the vapor. However, strong dipole moments lead to a positive potential difference. These two contrasting behaviors can be understood based on the previous discussion. Of particular interest are the intermediate dipole moments and shifts where, due to the competing dipole orientations in the liquid and vapor regions, for a given dipole moment ($\mu =1.6$ for example), increasing the dipole shift can cause an inversion in the sign of $\Delta \psi^*$. However, the point where $\Delta\psi^*\approx0$ does not necessitate that $\psi^*(z)=0$. As shown in Figure \ref{fig:E_psi_z}d, $\psi^*$ exhibits a minimum for $\mu=1.6$ and $d=0.15$ despite having $\Delta\psi^*=0$. While we are not aware if any real molecule corresponds to such parameters, this minimum in $\psi^*$ and corresponding S-shaped electric field could provide preferential localization for chemical reactions at liquid–vapor interfaces, even between atoms with similar partial charges. In addition, the sign inversion of $\Delta\psi$ implies that asymmetry alone could lead to the adsorption of differently charged ions for differing $\mu$ and $d$. We leave the exploration of electrolyte systems for a future study.

These results confirm that molecular asymmetry is a crucial factor in controlling interfacial electrostatic properties, and can even contribute a significant electric field in the absence of any other effects.

\section{Conclusion}

In this study, we conducted molecular dynamics simulations of the shifted Stockmayer fluid model to study the interfacial properties of asymmetric polar fluids. We systematically varied both the dipole shift, $d$, and the dipole moment, $\mu$, to observe the effect of the molecular properties on the equilibrium interfacial profiles. Specifically, we reported profiles for the density, nematic order (2nd Legendre polynomial of orientation), polar order, electric field, and electrostatic potential. The density of the liquid phase was enhanced by the presence of a dipole shift, which we attributed to the strengthened electrostatic interactions between dipoles that lie closer to the surface of the spherical particles. For strong dipole moments, and therefore high liquid densities, we encountered a ferroelectric transition known to occur in the Stockmayer fluid model. Interestingly, a moderate shift in the dipole completely destroyed the ordering in the ferroelectric transition as a result of the most favorable conformation changing from the typical end-aligned (parallel). A more detailed study of the bulk liquid equilibrium properties is currently in preparation (Venkatachalam et al. 2025).

We found that the presence of even a small shift in the dipole could lead to significant interfacial polarization. In line with previous MD simulations, we found that the molecules at the Gibbs dividing surface prefer to lie parallel to the interface ($P_2<0$). In contrast to previous simulations, but in agreement with theoretical predictions from classical density functional theory, our simulations revealed that vapor molecules just outside the interface have a preference to point perpendicular to the interface ($P_2>0$) for sufficiently strong dipole moments. To make simple analytical predictions for the angular distribution functions in proximity to the interface, we utilized the energy due to image dipole interactions. We extended the traditional image dipole interaction to include the effect of a dipole shift and found excellent agreement between the theoretical predictions and the distributions collected from MD simulations. We found that the regular Stockmayer fluid produces $P_2(z)<0$ in the liquid, $P_2(z)>0$ in the vapor, and $m(z)=0$ everywhere. On the other hand, the angular distribution functions of a shifted Stockmayer fluid are skewed towards the liquid ($\cos\theta<0$). While $P_2(z)$ does not significantly change, $m(z)$ is consistently negative, indicating a preference for the molecules to point towards the liquid.

We computed the polarization density, electric field, and the electrostatic potential difference between the two phases. The sign of the electric field displayed an inversion as the strength of the dipole moment was increased with a constant dipole shift. Namely, the electric field had the same sign as the density gradient for weak dipole moments and the opposite sign for strong dipole moments. At parameters consistent with those of water, we found that the interfacial electric field was roughly 16 MV/cm. We note that this agrees in both sign and magnitude with experimental observations and quantum mechanical MD simulations of water. The quantitative agreement is surprising and should not be read into since the shifted Stockmayer fluid is a minimal model that ignores many unique aspects of real fluids such as the mean inner potential, geometric asymmetry, polarization, and hydrogen bonding.

With this work, we have shown that slight molecular asymmetry in the Stockmayer fluid leads to rich interfacial behavior, including significant interfacial polarization. In the modern discussion of how interfacial electric fields catalyze reactions, this work provides context for a potentially important effect that is likely to be present to some extent in all polar fluids. In future studies we will explore the role of dipole shift on the bulk and interfacial properties of electrolyte solutions.

\begin{acknowledgments}

The authors would like to thank Dr. B. Ye for helping develop much of the tools required for the analysis used in this work. S.V. is supported by the U.S. Department of Energy, Office of Science, Office of Advanced Scientific Computing Research, Department of Energy Computational Science Graduate Fellowship under Award Number DE-SC0022158. Partial support for this research is provided by Hong Kong Quantum AI Lab, AIR@InnoHK of Hong Kong Government. B.Z. gratefully acknowledges support by Harvey Mudd College start-up grant and NSF CAREER Award CHE-2337602. Authors would like to thank the Resnick Sustainability Institute at Caltech for supporting A.V.’s research.

\end{acknowledgments}

\section*{Data Availability Statement}
The data that support the findings of this study are available from the corresponding author upon reasonable request.

\section*{References}
\bibliography{references,extra,ovito-ref}

\providecommand*{\mcitethebibliography}{\thebibliography}
\csname @ifundefined\endcsname{endmcitethebibliography}
{\let\endmcitethebibliography\endthebibliography}{}
\begin{mcitethebibliography}{1}
\providecommand*{\natexlab}[1]{#1}
\providecommand*{\mciteSetBstSublistMode}[1]{}
\providecommand*{\mciteSetBstMaxWidthForm}[2]{}
\providecommand*{\mciteBstWouldAddEndPuncttrue}
  {\def\EndOfBibitem{\unskip.}}
\providecommand*{\mciteBstWouldAddEndPunctfalse}
  {\let\EndOfBibitem\relax}
\providecommand*{\mciteSetBstMidEndSepPunct}[3]{}
\providecommand*{\mciteSetBstSublistLabelBeginEnd}[3]{}
\providecommand*{\EndOfBibitem}{}
\mciteSetBstSublistMode{f}
\mciteSetBstMaxWidthForm{subitem}
{(\emph{\alph{mcitesubitemcount}})}
\mciteSetBstSublistLabelBeginEnd{\mcitemaxwidthsubitemform\space}
{\relax}{\relax}

\bibitem[Willard and Chandler(2010)]{willard_instantaneous_2010}
A.~P. Willard and D.~Chandler, \emph{J. Phys. Chem. B}, 2010, \textbf{114}, 1954--1958\relax
\mciteBstWouldAddEndPuncttrue
\mciteSetBstMidEndSepPunct{\mcitedefaultmidpunct}
{\mcitedefaultendpunct}{\mcitedefaultseppunct}\relax
\EndOfBibitem
\end{mcitethebibliography}


%merlin.mbs apsrev4-1.bst 2010-07-25 4.21a (PWD, AO, DPC) hacked
%Control: key (0)
%Control: author (72) initials jnrlst
%Control: editor formatted (1) identically to author
%Control: production of article title (-1) disabled
%Control: page (0) single
%Control: year (1) truncated
%Control: production of eprint (0) enabled
\begin{thebibliography}{95}%
\makeatletter
\providecommand \@ifxundefined [1]{%
 \@ifx{#1\undefined}
}%
\providecommand \@ifnum [1]{%
 \ifnum #1\expandafter \@firstoftwo
 \else \expandafter \@secondoftwo
 \fi
}%
\providecommand \@ifx [1]{%
 \ifx #1\expandafter \@firstoftwo
 \else \expandafter \@secondoftwo
 \fi
}%
\providecommand \natexlab [1]{#1}%
\providecommand \enquote  [1]{``#1''}%
\providecommand \bibnamefont  [1]{#1}%
\providecommand \bibfnamefont [1]{#1}%
\providecommand \citenamefont [1]{#1}%
\providecommand \href@noop [0]{\@secondoftwo}%
\providecommand \href [0]{\begingroup \@sanitize@url \@href}%
\providecommand \@href[1]{\@@startlink{#1}\@@href}%
\providecommand \@@href[1]{\endgroup#1\@@endlink}%
\providecommand \@sanitize@url [0]{\catcode `\\12\catcode `\$12\catcode `\&12\catcode `\#12\catcode `\^12\catcode `\_12\catcode `\%12\relax}%
\providecommand \@@startlink[1]{}%
\providecommand \@@endlink[0]{}%
\providecommand \url  [0]{\begingroup\@sanitize@url \@url }%
\providecommand \@url [1]{\endgroup\@href {#1}{\urlprefix }}%
\providecommand \urlprefix  [0]{URL }%
\providecommand \Eprint [0]{\href }%
\providecommand \doibase [0]{http://dx.doi.org/}%
\providecommand \selectlanguage [0]{\@gobble}%
\providecommand \bibinfo  [0]{\@secondoftwo}%
\providecommand \bibfield  [0]{\@secondoftwo}%
\providecommand \translation [1]{[#1]}%
\providecommand \BibitemOpen [0]{}%
\providecommand \bibitemStop [0]{}%
\providecommand \bibitemNoStop [0]{.\EOS\space}%
\providecommand \EOS [0]{\spacefactor3000\relax}%
\providecommand \BibitemShut  [1]{\csname bibitem#1\endcsname}%
\let\auto@bib@innerbib\@empty
%</preamble>
\bibitem [{\citenamefont {Yan}\ \emph {et~al.}(2016)\citenamefont {Yan}, \citenamefont {Bain},\ and\ \citenamefont {Cooks}}]{yan_organic_2016}%
  \BibitemOpen
  \bibfield  {author} {\bibinfo {author} {\bibfnamefont {X.}~\bibnamefont {Yan}}, \bibinfo {author} {\bibfnamefont {R.~M.}\ \bibnamefont {Bain}}, \ and\ \bibinfo {author} {\bibfnamefont {R.~G.}\ \bibnamefont {Cooks}},\ }\href {\doibase 10.1002/anie.201602270} {\bibfield  {journal} {\bibinfo  {journal} {Angewandte Chemie International Edition}\ }\textbf {\bibinfo {volume} {55}},\ \bibinfo {pages} {12960} (\bibinfo {year} {2016})}\BibitemShut {NoStop}%
\bibitem [{\citenamefont {Wei}\ \emph {et~al.}(2020)\citenamefont {Wei}, \citenamefont {Li}, \citenamefont {Cooks},\ and\ \citenamefont {Yan}}]{wei_accelerated_2020}%
  \BibitemOpen
  \bibfield  {author} {\bibinfo {author} {\bibfnamefont {Z.}~\bibnamefont {Wei}}, \bibinfo {author} {\bibfnamefont {Y.}~\bibnamefont {Li}}, \bibinfo {author} {\bibfnamefont {R.~G.}\ \bibnamefont {Cooks}}, \ and\ \bibinfo {author} {\bibfnamefont {X.}~\bibnamefont {Yan}},\ }\href {\doibase 10.1146/annurev-physchem-121319-110654} {\bibfield  {journal} {\bibinfo  {journal} {Annual Review of Physical Chemistry}\ }\textbf {\bibinfo {volume} {71}},\ \bibinfo {pages} {31} (\bibinfo {year} {2020})}\BibitemShut {NoStop}%
\bibitem [{\citenamefont {Narayan}\ \emph {et~al.}(2005)\citenamefont {Narayan}, \citenamefont {Muldoon}, \citenamefont {Finn}, \citenamefont {Fokin}, \citenamefont {Kolb},\ and\ \citenamefont {Sharpless}}]{narayan_water_2005}%
  \BibitemOpen
  \bibfield  {author} {\bibinfo {author} {\bibfnamefont {S.}~\bibnamefont {Narayan}}, \bibinfo {author} {\bibfnamefont {J.}~\bibnamefont {Muldoon}}, \bibinfo {author} {\bibfnamefont {M.~G.}\ \bibnamefont {Finn}}, \bibinfo {author} {\bibfnamefont {V.~V.}\ \bibnamefont {Fokin}}, \bibinfo {author} {\bibfnamefont {H.~C.}\ \bibnamefont {Kolb}}, \ and\ \bibinfo {author} {\bibfnamefont {K.~B.}\ \bibnamefont {Sharpless}},\ }\href {\doibase 10.1002/anie.200462883} {\bibfield  {journal} {\bibinfo  {journal} {Angewandte Chemie International Edition}\ }\textbf {\bibinfo {volume} {44}},\ \bibinfo {pages} {3275} (\bibinfo {year} {2005})}\BibitemShut {NoStop}%
\bibitem [{\citenamefont {Serrano-Luginbühl}\ \emph {et~al.}(2018)\citenamefont {Serrano-Luginbühl}, \citenamefont {Ruiz-Mirazo}, \citenamefont {Ostaszewski}, \citenamefont {Gallou},\ and\ \citenamefont {Walde}}]{serrano-luginbuhl_soft_2018}%
  \BibitemOpen
  \bibfield  {author} {\bibinfo {author} {\bibfnamefont {S.}~\bibnamefont {Serrano-Luginbühl}}, \bibinfo {author} {\bibfnamefont {K.}~\bibnamefont {Ruiz-Mirazo}}, \bibinfo {author} {\bibfnamefont {R.}~\bibnamefont {Ostaszewski}}, \bibinfo {author} {\bibfnamefont {F.}~\bibnamefont {Gallou}}, \ and\ \bibinfo {author} {\bibfnamefont {P.}~\bibnamefont {Walde}},\ }\href {\doibase 10.1038/s41570-018-0042-6} {\bibfield  {journal} {\bibinfo  {journal} {Nat Rev Chem}\ }\textbf {\bibinfo {volume} {2}},\ \bibinfo {pages} {306} (\bibinfo {year} {2018})}\BibitemShut {NoStop}%
\bibitem [{\citenamefont {Jordan}\ \emph {et~al.}(2022)\citenamefont {Jordan}, \citenamefont {Lowe},\ and\ \citenamefont {Verlet}}]{jordan_photooxidation_2022}%
  \BibitemOpen
  \bibfield  {author} {\bibinfo {author} {\bibfnamefont {C.~J.~C.}\ \bibnamefont {Jordan}}, \bibinfo {author} {\bibfnamefont {E.~A.}\ \bibnamefont {Lowe}}, \ and\ \bibinfo {author} {\bibfnamefont {J.~R.~R.}\ \bibnamefont {Verlet}},\ }\href {\doibase 10.1021/jacs.2c04935} {\bibfield  {journal} {\bibinfo  {journal} {J. Am. Chem. Soc.}\ }\textbf {\bibinfo {volume} {144}},\ \bibinfo {pages} {14012} (\bibinfo {year} {2022})}\BibitemShut {NoStop}%
\bibitem [{\citenamefont {Chen}\ \emph {et~al.}(2023)\citenamefont {Chen}, \citenamefont {Wang}, \citenamefont {Xu}, \citenamefont {Yuan}, \citenamefont {Zhang}, \citenamefont {Zhu}, \citenamefont {Marshall}, \citenamefont {Bowen},\ and\ \citenamefont {Zhang}}]{chen_spontaneous_2023}%
  \BibitemOpen
  \bibfield  {author} {\bibinfo {author} {\bibfnamefont {H.}~\bibnamefont {Chen}}, \bibinfo {author} {\bibfnamefont {R.}~\bibnamefont {Wang}}, \bibinfo {author} {\bibfnamefont {J.}~\bibnamefont {Xu}}, \bibinfo {author} {\bibfnamefont {X.}~\bibnamefont {Yuan}}, \bibinfo {author} {\bibfnamefont {D.}~\bibnamefont {Zhang}}, \bibinfo {author} {\bibfnamefont {Z.}~\bibnamefont {Zhu}}, \bibinfo {author} {\bibfnamefont {M.}~\bibnamefont {Marshall}}, \bibinfo {author} {\bibfnamefont {K.}~\bibnamefont {Bowen}}, \ and\ \bibinfo {author} {\bibfnamefont {X.}~\bibnamefont {Zhang}},\ }\href {\doibase 10.1021/jacs.2c12731} {\bibfield  {journal} {\bibinfo  {journal} {J. Am. Chem. Soc.}\ }\textbf {\bibinfo {volume} {145}},\ \bibinfo {pages} {2647} (\bibinfo {year} {2023})}\BibitemShut {NoStop}%
\bibitem [{\citenamefont {Liu}\ and\ \citenamefont {Abbatt}(2021)}]{liu_oxidation_2021}%
  \BibitemOpen
  \bibfield  {author} {\bibinfo {author} {\bibfnamefont {T.}~\bibnamefont {Liu}}\ and\ \bibinfo {author} {\bibfnamefont {J.~P.~D.}\ \bibnamefont {Abbatt}},\ }\href {\doibase 10.1038/s41557-021-00777-0} {\bibfield  {journal} {\bibinfo  {journal} {Nat. Chem.}\ }\textbf {\bibinfo {volume} {13}},\ \bibinfo {pages} {1173} (\bibinfo {year} {2021})}\BibitemShut {NoStop}%
\bibitem [{\citenamefont {Wan}\ \emph {et~al.}(2023)\citenamefont {Wan}, \citenamefont {Fang}, \citenamefont {Liu}, \citenamefont {Francisco},\ and\ \citenamefont {Zhu}}]{wan_mechanistic_2023}%
  \BibitemOpen
  \bibfield  {author} {\bibinfo {author} {\bibfnamefont {Z.}~\bibnamefont {Wan}}, \bibinfo {author} {\bibfnamefont {Y.}~\bibnamefont {Fang}}, \bibinfo {author} {\bibfnamefont {Z.}~\bibnamefont {Liu}}, \bibinfo {author} {\bibfnamefont {J.~S.}\ \bibnamefont {Francisco}}, \ and\ \bibinfo {author} {\bibfnamefont {C.}~\bibnamefont {Zhu}},\ }\href {\doibase 10.1021/jacs.2c09837} {\bibfield  {journal} {\bibinfo  {journal} {J. Am. Chem. Soc.}\ }\textbf {\bibinfo {volume} {145}},\ \bibinfo {pages} {944} (\bibinfo {year} {2023})}\BibitemShut {NoStop}%
\bibitem [{\citenamefont {Rossignol}\ \emph {et~al.}(2016)\citenamefont {Rossignol}, \citenamefont {Tinel}, \citenamefont {Bianco}, \citenamefont {Passananti}, \citenamefont {Brigante}, \citenamefont {Donaldson},\ and\ \citenamefont {George}}]{rossignol_atmospheric_2016}%
  \BibitemOpen
  \bibfield  {author} {\bibinfo {author} {\bibfnamefont {S.}~\bibnamefont {Rossignol}}, \bibinfo {author} {\bibfnamefont {L.}~\bibnamefont {Tinel}}, \bibinfo {author} {\bibfnamefont {A.}~\bibnamefont {Bianco}}, \bibinfo {author} {\bibfnamefont {M.}~\bibnamefont {Passananti}}, \bibinfo {author} {\bibfnamefont {M.}~\bibnamefont {Brigante}}, \bibinfo {author} {\bibfnamefont {D.~J.}\ \bibnamefont {Donaldson}}, \ and\ \bibinfo {author} {\bibfnamefont {C.}~\bibnamefont {George}},\ }\href {\doibase 10.1126/science.aaf3617} {\bibfield  {journal} {\bibinfo  {journal} {Science}\ }\textbf {\bibinfo {volume} {353}},\ \bibinfo {pages} {699} (\bibinfo {year} {2016})}\BibitemShut {NoStop}%
\bibitem [{\citenamefont {Xia}\ \emph {et~al.}(2023)\citenamefont {Xia}, \citenamefont {Chen}, \citenamefont {Xie}, \citenamefont {Zhong},\ and\ \citenamefont {Francisco}}]{xia_counterintuitive_2023}%
  \BibitemOpen
  \bibfield  {author} {\bibinfo {author} {\bibfnamefont {D.}~\bibnamefont {Xia}}, \bibinfo {author} {\bibfnamefont {J.}~\bibnamefont {Chen}}, \bibinfo {author} {\bibfnamefont {H.-B.}\ \bibnamefont {Xie}}, \bibinfo {author} {\bibfnamefont {J.}~\bibnamefont {Zhong}}, \ and\ \bibinfo {author} {\bibfnamefont {J.~S.}\ \bibnamefont {Francisco}},\ }\href {\doibase 10.1021/jacs.2c13661} {\bibfield  {journal} {\bibinfo  {journal} {J. Am. Chem. Soc.}\ }\textbf {\bibinfo {volume} {145}},\ \bibinfo {pages} {4791} (\bibinfo {year} {2023})}\BibitemShut {NoStop}%
\bibitem [{\citenamefont {Liang}\ \emph {et~al.}(2023)\citenamefont {Liang}, \citenamefont {Zhu},\ and\ \citenamefont {Yang}}]{liang_water_2023}%
  \BibitemOpen
  \bibfield  {author} {\bibinfo {author} {\bibfnamefont {Q.}~\bibnamefont {Liang}}, \bibinfo {author} {\bibfnamefont {C.}~\bibnamefont {Zhu}}, \ and\ \bibinfo {author} {\bibfnamefont {J.}~\bibnamefont {Yang}},\ }\href {\doibase 10.1021/jacs.3c00734} {\bibfield  {journal} {\bibinfo  {journal} {J. Am. Chem. Soc.}\ }\textbf {\bibinfo {volume} {145}},\ \bibinfo {pages} {10159} (\bibinfo {year} {2023})}\BibitemShut {NoStop}%
\bibitem [{\citenamefont {George}\ \emph {et~al.}(2015)\citenamefont {George}, \citenamefont {Ammann}, \citenamefont {D’Anna}, \citenamefont {Donaldson},\ and\ \citenamefont {Nizkorodov}}]{george_heterogeneous_2015}%
  \BibitemOpen
  \bibfield  {author} {\bibinfo {author} {\bibfnamefont {C.}~\bibnamefont {George}}, \bibinfo {author} {\bibfnamefont {M.}~\bibnamefont {Ammann}}, \bibinfo {author} {\bibfnamefont {B.}~\bibnamefont {D’Anna}}, \bibinfo {author} {\bibfnamefont {D.~J.}\ \bibnamefont {Donaldson}}, \ and\ \bibinfo {author} {\bibfnamefont {S.~A.}\ \bibnamefont {Nizkorodov}},\ }\href {\doibase 10.1021/cr500648z} {\bibfield  {journal} {\bibinfo  {journal} {Chem. Rev.}\ }\textbf {\bibinfo {volume} {115}},\ \bibinfo {pages} {4218} (\bibinfo {year} {2015})}\BibitemShut {NoStop}%
\bibitem [{\citenamefont {Ruiz-Lopez}\ \emph {et~al.}(2020)\citenamefont {Ruiz-Lopez}, \citenamefont {Francisco}, \citenamefont {Martins-Costa},\ and\ \citenamefont {Anglada}}]{ruiz-lopez_molecular_2020}%
  \BibitemOpen
  \bibfield  {author} {\bibinfo {author} {\bibfnamefont {M.~F.}\ \bibnamefont {Ruiz-Lopez}}, \bibinfo {author} {\bibfnamefont {J.~S.}\ \bibnamefont {Francisco}}, \bibinfo {author} {\bibfnamefont {M.~T.~C.}\ \bibnamefont {Martins-Costa}}, \ and\ \bibinfo {author} {\bibfnamefont {J.~M.}\ \bibnamefont {Anglada}},\ }\href {\doibase 10.1038/s41570-020-0203-2} {\bibfield  {journal} {\bibinfo  {journal} {Nat Rev Chem}\ }\textbf {\bibinfo {volume} {4}},\ \bibinfo {pages} {459} (\bibinfo {year} {2020})}\BibitemShut {NoStop}%
\bibitem [{\citenamefont {Yan}\ \emph {et~al.}(2013)\citenamefont {Yan}, \citenamefont {Augusti}, \citenamefont {Li},\ and\ \citenamefont {Cooks}}]{yan_chemical_2013}%
  \BibitemOpen
  \bibfield  {author} {\bibinfo {author} {\bibfnamefont {X.}~\bibnamefont {Yan}}, \bibinfo {author} {\bibfnamefont {R.}~\bibnamefont {Augusti}}, \bibinfo {author} {\bibfnamefont {X.}~\bibnamefont {Li}}, \ and\ \bibinfo {author} {\bibfnamefont {R.~G.}\ \bibnamefont {Cooks}},\ }\href {\doibase 10.1002/cplu.201300172} {\bibfield  {journal} {\bibinfo  {journal} {ChemPlusChem}\ }\textbf {\bibinfo {volume} {78}},\ \bibinfo {pages} {1142} (\bibinfo {year} {2013})}\BibitemShut {NoStop}%
\bibitem [{\citenamefont {Girod}\ \emph {et~al.}(2011)\citenamefont {Girod}, \citenamefont {Moyano}, \citenamefont {Campbell},\ and\ \citenamefont {Cooks}}]{girod_accelerated_2011}%
  \BibitemOpen
  \bibfield  {author} {\bibinfo {author} {\bibfnamefont {M.}~\bibnamefont {Girod}}, \bibinfo {author} {\bibfnamefont {E.}~\bibnamefont {Moyano}}, \bibinfo {author} {\bibfnamefont {D.~I.}\ \bibnamefont {Campbell}}, \ and\ \bibinfo {author} {\bibfnamefont {R.~G.}\ \bibnamefont {Cooks}},\ }\href {\doibase 10.1039/C0SC00416B} {\bibfield  {journal} {\bibinfo  {journal} {Chem. Sci.}\ }\textbf {\bibinfo {volume} {2}},\ \bibinfo {pages} {501} (\bibinfo {year} {2011})}\BibitemShut {NoStop}%
\bibitem [{\citenamefont {Pestana}\ \emph {et~al.}(2020)\citenamefont {Pestana}, \citenamefont {Hao},\ and\ \citenamefont {Head-Gordon}}]{pestana_dielsalder_2020}%
  \BibitemOpen
  \bibfield  {author} {\bibinfo {author} {\bibfnamefont {L.~R.}\ \bibnamefont {Pestana}}, \bibinfo {author} {\bibfnamefont {H.}~\bibnamefont {Hao}}, \ and\ \bibinfo {author} {\bibfnamefont {T.}~\bibnamefont {Head-Gordon}},\ }\href {\doibase 10.1021/acs.nanolett.9b04369} {\bibfield  {journal} {\bibinfo  {journal} {Nano Lett.}\ }\textbf {\bibinfo {volume} {20}},\ \bibinfo {pages} {606} (\bibinfo {year} {2020})}\BibitemShut {NoStop}%
\bibitem [{\citenamefont {Matyushov}(2019)}]{matyushov_electrostatic_2019}%
  \BibitemOpen
  \bibfield  {author} {\bibinfo {author} {\bibfnamefont {D.~V.}\ \bibnamefont {Matyushov}},\ }\href {\doibase 10.1063/1.5124390} {\bibfield  {journal} {\bibinfo  {journal} {Biomicrofluidics}\ }\textbf {\bibinfo {volume} {13}},\ \bibinfo {pages} {064106} (\bibinfo {year} {2019})}\BibitemShut {NoStop}%
\bibitem [{\citenamefont {Muñoz-Santiburcio}\ and\ \citenamefont {Marx}(2021)}]{munoz-santiburcio_confinement-controlled_2021}%
  \BibitemOpen
  \bibfield  {author} {\bibinfo {author} {\bibfnamefont {D.}~\bibnamefont {Muñoz-Santiburcio}}\ and\ \bibinfo {author} {\bibfnamefont {D.}~\bibnamefont {Marx}},\ }\href {\doibase 10.1021/acs.chemrev.0c01292} {\bibfield  {journal} {\bibinfo  {journal} {Chem. Rev.}\ }\textbf {\bibinfo {volume} {121}},\ \bibinfo {pages} {6293} (\bibinfo {year} {2021})}\BibitemShut {NoStop}%
\bibitem [{\citenamefont {Welborn}\ \emph {et~al.}(2018)\citenamefont {Welborn}, \citenamefont {Ruiz~Pestana},\ and\ \citenamefont {Head-Gordon}}]{welborn_computational_2018}%
  \BibitemOpen
  \bibfield  {author} {\bibinfo {author} {\bibfnamefont {V.~V.}\ \bibnamefont {Welborn}}, \bibinfo {author} {\bibfnamefont {L.}~\bibnamefont {Ruiz~Pestana}}, \ and\ \bibinfo {author} {\bibfnamefont {T.}~\bibnamefont {Head-Gordon}},\ }\href {\doibase 10.1038/s41929-018-0109-2} {\bibfield  {journal} {\bibinfo  {journal} {Nat Catal}\ }\textbf {\bibinfo {volume} {1}},\ \bibinfo {pages} {649} (\bibinfo {year} {2018})}\BibitemShut {NoStop}%
\bibitem [{\citenamefont {Ashton}\ \emph {et~al.}(2020)\citenamefont {Ashton}, \citenamefont {Mishra}, \citenamefont {Neugebauer},\ and\ \citenamefont {Freysoldt}}]{ashton_ab_2020}%
  \BibitemOpen
  \bibfield  {author} {\bibinfo {author} {\bibfnamefont {M.}~\bibnamefont {Ashton}}, \bibinfo {author} {\bibfnamefont {A.}~\bibnamefont {Mishra}}, \bibinfo {author} {\bibfnamefont {J.}~\bibnamefont {Neugebauer}}, \ and\ \bibinfo {author} {\bibfnamefont {C.}~\bibnamefont {Freysoldt}},\ }\href {\doibase 10.1103/PhysRevLett.124.176801} {\bibfield  {journal} {\bibinfo  {journal} {Phys. Rev. Lett.}\ }\textbf {\bibinfo {volume} {124}},\ \bibinfo {pages} {176801} (\bibinfo {year} {2020})}\BibitemShut {NoStop}%
\bibitem [{\citenamefont {Shaik}\ \emph {et~al.}(2020)\citenamefont {Shaik}, \citenamefont {Danovich}, \citenamefont {Joy}, \citenamefont {Wang},\ and\ \citenamefont {Stuyver}}]{shaik_electric-field_2020}%
  \BibitemOpen
  \bibfield  {author} {\bibinfo {author} {\bibfnamefont {S.}~\bibnamefont {Shaik}}, \bibinfo {author} {\bibfnamefont {D.}~\bibnamefont {Danovich}}, \bibinfo {author} {\bibfnamefont {J.}~\bibnamefont {Joy}}, \bibinfo {author} {\bibfnamefont {Z.}~\bibnamefont {Wang}}, \ and\ \bibinfo {author} {\bibfnamefont {T.}~\bibnamefont {Stuyver}},\ }\href {\doibase 10.1021/jacs.0c05128} {\bibfield  {journal} {\bibinfo  {journal} {J. Am. Chem. Soc.}\ }\textbf {\bibinfo {volume} {142}},\ \bibinfo {pages} {12551} (\bibinfo {year} {2020})}\BibitemShut {NoStop}%
\bibitem [{\citenamefont {Bím}\ and\ \citenamefont {Alexandrova}(2021)}]{bim_local_2021}%
  \BibitemOpen
  \bibfield  {author} {\bibinfo {author} {\bibfnamefont {D.}~\bibnamefont {Bím}}\ and\ \bibinfo {author} {\bibfnamefont {A.~N.}\ \bibnamefont {Alexandrova}},\ }\href {\doibase 10.1021/acscatal.1c00687} {\bibfield  {journal} {\bibinfo  {journal} {ACS Catal.}\ }\textbf {\bibinfo {volume} {11}},\ \bibinfo {pages} {6534} (\bibinfo {year} {2021})}\BibitemShut {NoStop}%
\bibitem [{\citenamefont {Eberhart}\ \emph {et~al.}(2024)\citenamefont {Eberhart}, \citenamefont {Wilson}, \citenamefont {Jones},\ and\ \citenamefont {Alexandrova}}]{eberhart_electric_2024}%
  \BibitemOpen
  \bibfield  {author} {\bibinfo {author} {\bibfnamefont {M.~E.}\ \bibnamefont {Eberhart}}, \bibinfo {author} {\bibfnamefont {T.~R.}\ \bibnamefont {Wilson}}, \bibinfo {author} {\bibfnamefont {T.~E.}\ \bibnamefont {Jones}}, \ and\ \bibinfo {author} {\bibfnamefont {A.~N.}\ \bibnamefont {Alexandrova}},\ }\href {\doibase 10.1073/pnas.2411976121} {\bibfield  {journal} {\bibinfo  {journal} {Proceedings of the National Academy of Sciences}\ }\textbf {\bibinfo {volume} {121}},\ \bibinfo {pages} {e2411976121} (\bibinfo {year} {2024})}\BibitemShut {NoStop}%
\bibitem [{\citenamefont {Aragonès}\ \emph {et~al.}(2016)\citenamefont {Aragonès}, \citenamefont {Haworth}, \citenamefont {Darwish}, \citenamefont {Ciampi}, \citenamefont {Mannix}, \citenamefont {Wallace}, \citenamefont {Diez-Perez},\ and\ \citenamefont {Coote}}]{aragones_electrostatic_2016}%
  \BibitemOpen
  \bibfield  {author} {\bibinfo {author} {\bibfnamefont {A.~C.}\ \bibnamefont {Aragonès}}, \bibinfo {author} {\bibfnamefont {N.~L.}\ \bibnamefont {Haworth}}, \bibinfo {author} {\bibfnamefont {N.}~\bibnamefont {Darwish}}, \bibinfo {author} {\bibfnamefont {S.}~\bibnamefont {Ciampi}}, \bibinfo {author} {\bibfnamefont {E.~J.}\ \bibnamefont {Mannix}}, \bibinfo {author} {\bibfnamefont {G.~G.}\ \bibnamefont {Wallace}}, \bibinfo {author} {\bibfnamefont {I.}~\bibnamefont {Diez-Perez}}, \ and\ \bibinfo {author} {\bibfnamefont {M.~L.}\ \bibnamefont {Coote}},\ }\href {\doibase 10.1038/nature16989} {\bibfield  {journal} {\bibinfo  {journal} {Nature}\ }\textbf {\bibinfo {volume} {531}},\ \bibinfo {pages} {88} (\bibinfo {year} {2016})}\BibitemShut {NoStop}%
\bibitem [{\citenamefont {Shaik}\ \emph {et~al.}(2016)\citenamefont {Shaik}, \citenamefont {Mandal},\ and\ \citenamefont {Ramanan}}]{shaik_oriented_2016}%
  \BibitemOpen
  \bibfield  {author} {\bibinfo {author} {\bibfnamefont {S.}~\bibnamefont {Shaik}}, \bibinfo {author} {\bibfnamefont {D.}~\bibnamefont {Mandal}}, \ and\ \bibinfo {author} {\bibfnamefont {R.}~\bibnamefont {Ramanan}},\ }\href {\doibase 10.1038/nchem.2651} {\bibfield  {journal} {\bibinfo  {journal} {Nature Chem}\ }\textbf {\bibinfo {volume} {8}},\ \bibinfo {pages} {1091} (\bibinfo {year} {2016})}\BibitemShut {NoStop}%
\bibitem [{\citenamefont {Stuyver}\ \emph {et~al.}(2020)\citenamefont {Stuyver}, \citenamefont {Danovich}, \citenamefont {Joy},\ and\ \citenamefont {Shaik}}]{stuyver_external_2020}%
  \BibitemOpen
  \bibfield  {author} {\bibinfo {author} {\bibfnamefont {T.}~\bibnamefont {Stuyver}}, \bibinfo {author} {\bibfnamefont {D.}~\bibnamefont {Danovich}}, \bibinfo {author} {\bibfnamefont {J.}~\bibnamefont {Joy}}, \ and\ \bibinfo {author} {\bibfnamefont {S.}~\bibnamefont {Shaik}},\ }\href {\doibase 10.1002/wcms.1438} {\bibfield  {journal} {\bibinfo  {journal} {WIREs Computational Molecular Science}\ }\textbf {\bibinfo {volume} {10}},\ \bibinfo {pages} {e1438} (\bibinfo {year} {2020})}\BibitemShut {NoStop}%
\bibitem [{\citenamefont {Zheng}\ \emph {et~al.}(2022)\citenamefont {Zheng}, \citenamefont {Mao}, \citenamefont {Kozuch}, \citenamefont {Atsango}, \citenamefont {Ji}, \citenamefont {Markland},\ and\ \citenamefont {Boxer}}]{zheng_two-directional_2022}%
  \BibitemOpen
  \bibfield  {author} {\bibinfo {author} {\bibfnamefont {C.}~\bibnamefont {Zheng}}, \bibinfo {author} {\bibfnamefont {Y.}~\bibnamefont {Mao}}, \bibinfo {author} {\bibfnamefont {J.}~\bibnamefont {Kozuch}}, \bibinfo {author} {\bibfnamefont {A.~O.}\ \bibnamefont {Atsango}}, \bibinfo {author} {\bibfnamefont {Z.}~\bibnamefont {Ji}}, \bibinfo {author} {\bibfnamefont {T.~E.}\ \bibnamefont {Markland}}, \ and\ \bibinfo {author} {\bibfnamefont {S.~G.}\ \bibnamefont {Boxer}},\ }\href {\doibase 10.1038/s41557-022-00937-w} {\bibfield  {journal} {\bibinfo  {journal} {Nat. Chem.}\ }\textbf {\bibinfo {volume} {14}},\ \bibinfo {pages} {891} (\bibinfo {year} {2022})}\BibitemShut {NoStop}%
\bibitem [{\citenamefont {Welborn}\ and\ \citenamefont {Head-Gordon}(2019)}]{welborn_fluctuations_2019}%
  \BibitemOpen
  \bibfield  {author} {\bibinfo {author} {\bibfnamefont {V.~V.}\ \bibnamefont {Welborn}}\ and\ \bibinfo {author} {\bibfnamefont {T.}~\bibnamefont {Head-Gordon}},\ }\href {\doibase 10.1021/jacs.9b05323} {\bibfield  {journal} {\bibinfo  {journal} {J. Am. Chem. Soc.}\ }\textbf {\bibinfo {volume} {141}},\ \bibinfo {pages} {12487} (\bibinfo {year} {2019})}\BibitemShut {NoStop}%
\bibitem [{\citenamefont {Morita}\ and\ \citenamefont {Hynes}(2000)}]{morita_theoretical_2000}%
  \BibitemOpen
  \bibfield  {author} {\bibinfo {author} {\bibfnamefont {A.}~\bibnamefont {Morita}}\ and\ \bibinfo {author} {\bibfnamefont {J.~T.}\ \bibnamefont {Hynes}},\ }\href {\doibase 10.1016/S0301-0104(00)00127-0} {\bibfield  {journal} {\bibinfo  {journal} {Chemical Physics}\ }\textbf {\bibinfo {volume} {258}},\ \bibinfo {pages} {371} (\bibinfo {year} {2000})}\BibitemShut {NoStop}%
\bibitem [{\citenamefont {Du}\ \emph {et~al.}(1993)\citenamefont {Du}, \citenamefont {Superfine}, \citenamefont {Freysz},\ and\ \citenamefont {Shen}}]{du_vibrational_1993}%
  \BibitemOpen
  \bibfield  {author} {\bibinfo {author} {\bibfnamefont {Q.}~\bibnamefont {Du}}, \bibinfo {author} {\bibfnamefont {R.}~\bibnamefont {Superfine}}, \bibinfo {author} {\bibfnamefont {E.}~\bibnamefont {Freysz}}, \ and\ \bibinfo {author} {\bibfnamefont {Y.~R.}\ \bibnamefont {Shen}},\ }\href {\doibase 10.1103/PhysRevLett.70.2313} {\bibfield  {journal} {\bibinfo  {journal} {Phys. Rev. Lett.}\ }\textbf {\bibinfo {volume} {70}},\ \bibinfo {pages} {2313} (\bibinfo {year} {1993})}\BibitemShut {NoStop}%
\bibitem [{\citenamefont {Medders}\ and\ \citenamefont {Paesani}(2016)}]{medders_dissecting_2016}%
  \BibitemOpen
  \bibfield  {author} {\bibinfo {author} {\bibfnamefont {G.~R.}\ \bibnamefont {Medders}}\ and\ \bibinfo {author} {\bibfnamefont {F.}~\bibnamefont {Paesani}},\ }\href {\doibase 10.1021/jacs.6b00893} {\bibfield  {journal} {\bibinfo  {journal} {J. Am. Chem. Soc.}\ }\textbf {\bibinfo {volume} {138}},\ \bibinfo {pages} {3912} (\bibinfo {year} {2016})}\BibitemShut {NoStop}%
\bibitem [{\citenamefont {Moberg}\ \emph {et~al.}(2018)\citenamefont {Moberg}, \citenamefont {Straight},\ and\ \citenamefont {Paesani}}]{moberg_temperature_2018}%
  \BibitemOpen
  \bibfield  {author} {\bibinfo {author} {\bibfnamefont {D.~R.}\ \bibnamefont {Moberg}}, \bibinfo {author} {\bibfnamefont {S.~C.}\ \bibnamefont {Straight}}, \ and\ \bibinfo {author} {\bibfnamefont {F.}~\bibnamefont {Paesani}},\ }\href {\doibase 10.1021/acs.jpcb.8b01726} {\bibfield  {journal} {\bibinfo  {journal} {J. Phys. Chem. B}\ }\textbf {\bibinfo {volume} {122}},\ \bibinfo {pages} {4356} (\bibinfo {year} {2018})}\BibitemShut {NoStop}%
\bibitem [{\citenamefont {Pezzotti}\ \emph {et~al.}(2017)\citenamefont {Pezzotti}, \citenamefont {Galimberti},\ and\ \citenamefont {Gaigeot}}]{pezzotti_2d_2017}%
  \BibitemOpen
  \bibfield  {author} {\bibinfo {author} {\bibfnamefont {S.}~\bibnamefont {Pezzotti}}, \bibinfo {author} {\bibfnamefont {D.~R.}\ \bibnamefont {Galimberti}}, \ and\ \bibinfo {author} {\bibfnamefont {M.-P.}\ \bibnamefont {Gaigeot}},\ }\href {\doibase 10.1021/acs.jpclett.7b01257} {\bibfield  {journal} {\bibinfo  {journal} {J. Phys. Chem. Lett.}\ }\textbf {\bibinfo {volume} {8}},\ \bibinfo {pages} {3133} (\bibinfo {year} {2017})}\BibitemShut {NoStop}%
\bibitem [{\citenamefont {Pezzotti}\ \emph {et~al.}(2018)\citenamefont {Pezzotti}, \citenamefont {Serva},\ and\ \citenamefont {Gaigeot}}]{pezzotti_2d-hb-network_2018}%
  \BibitemOpen
  \bibfield  {author} {\bibinfo {author} {\bibfnamefont {S.}~\bibnamefont {Pezzotti}}, \bibinfo {author} {\bibfnamefont {A.}~\bibnamefont {Serva}}, \ and\ \bibinfo {author} {\bibfnamefont {M.-P.}\ \bibnamefont {Gaigeot}},\ }\href {\doibase 10.1063/1.5018096} {\bibfield  {journal} {\bibinfo  {journal} {J. Chem. Phys.}\ }\textbf {\bibinfo {volume} {148}},\ \bibinfo {pages} {174701} (\bibinfo {year} {2018})}\BibitemShut {NoStop}%
\bibitem [{\citenamefont {Cooper}\ \emph {et~al.}(2017)\citenamefont {Cooper}, \citenamefont {O'Brien}, \citenamefont {Chang},\ and\ \citenamefont {Williams}}]{cooper_structural_2017}%
  \BibitemOpen
  \bibfield  {author} {\bibinfo {author} {\bibfnamefont {R.~J.}\ \bibnamefont {Cooper}}, \bibinfo {author} {\bibfnamefont {J.~T.}\ \bibnamefont {O'Brien}}, \bibinfo {author} {\bibfnamefont {T.~M.}\ \bibnamefont {Chang}}, \ and\ \bibinfo {author} {\bibfnamefont {E.~R.}\ \bibnamefont {Williams}},\ }\href {\doibase 10.1039/C7SC00481H} {\bibfield  {journal} {\bibinfo  {journal} {Chem. Sci.}\ }\textbf {\bibinfo {volume} {8}},\ \bibinfo {pages} {5201} (\bibinfo {year} {2017})}\BibitemShut {NoStop}%
\bibitem [{\citenamefont {Hao}\ \emph {et~al.}(2022)\citenamefont {Hao}, \citenamefont {Leven},\ and\ \citenamefont {Head-Gordon}}]{hao_can_2022}%
  \BibitemOpen
  \bibfield  {author} {\bibinfo {author} {\bibfnamefont {H.}~\bibnamefont {Hao}}, \bibinfo {author} {\bibfnamefont {I.}~\bibnamefont {Leven}}, \ and\ \bibinfo {author} {\bibfnamefont {T.}~\bibnamefont {Head-Gordon}},\ }\href {\doibase 10.1038/s41467-021-27941-x} {\bibfield  {journal} {\bibinfo  {journal} {Nat Commun}\ }\textbf {\bibinfo {volume} {13}},\ \bibinfo {pages} {280} (\bibinfo {year} {2022})}\BibitemShut {NoStop}%
\bibitem [{\citenamefont {Kathmann}\ \emph {et~al.}(2011)\citenamefont {Kathmann}, \citenamefont {Kuo}, \citenamefont {Mundy},\ and\ \citenamefont {Schenter}}]{kathmann_understanding_2011}%
  \BibitemOpen
  \bibfield  {author} {\bibinfo {author} {\bibfnamefont {S.~M.}\ \bibnamefont {Kathmann}}, \bibinfo {author} {\bibfnamefont {I.-F.~W.}\ \bibnamefont {Kuo}}, \bibinfo {author} {\bibfnamefont {C.~J.}\ \bibnamefont {Mundy}}, \ and\ \bibinfo {author} {\bibfnamefont {G.~K.}\ \bibnamefont {Schenter}},\ }\href {\doibase 10.1021/jp1116036} {\bibfield  {journal} {\bibinfo  {journal} {J. Phys. Chem. B}\ }\textbf {\bibinfo {volume} {115}},\ \bibinfo {pages} {4369} (\bibinfo {year} {2011})}\BibitemShut {NoStop}%
\bibitem [{\citenamefont {Stillinger}(1980)}]{stillinger_water_1980}%
  \BibitemOpen
  \bibfield  {author} {\bibinfo {author} {\bibfnamefont {F.~H.}\ \bibnamefont {Stillinger}},\ }\href {\doibase 10.1126/science.209.4455.451} {\bibfield  {journal} {\bibinfo  {journal} {Science}\ }\textbf {\bibinfo {volume} {209}},\ \bibinfo {pages} {451} (\bibinfo {year} {1980})}\BibitemShut {NoStop}%
\bibitem [{\citenamefont {Franks}(2000)}]{franks_water_2000}%
  \BibitemOpen
  \bibfield  {author} {\bibinfo {author} {\bibfnamefont {F.}~\bibnamefont {Franks}},\ }\href@noop {} {{\selectlanguage {en}\emph {\bibinfo {title} {Water: {A} {Matrix} of {Life}}}}}\ (\bibinfo  {publisher} {Royal Society of Chemistry},\ \bibinfo {year} {2000})\BibitemShut {NoStop}%
\bibitem [{\citenamefont {Ball}(2001)}]{ball_lifes_2001}%
  \BibitemOpen
  \bibfield  {author} {\bibinfo {author} {\bibfnamefont {P.}~\bibnamefont {Ball}},\ }\href@noop {} {{\selectlanguage {en}\emph {\bibinfo {title} {Life's {Matrix}: {A} {Biography} of {Water}}}}}\ (\bibinfo  {publisher} {University of California Press},\ \bibinfo {year} {2001})\BibitemShut {NoStop}%
\bibitem [{\citenamefont {Stockmayer}(1941{\natexlab{a}})}]{stockmayer_second_1941}%
  \BibitemOpen
  \bibfield  {author} {\bibinfo {author} {\bibfnamefont {W.~H.}\ \bibnamefont {Stockmayer}},\ }\href {\doibase 10.1063/1.1750858} {\bibfield  {journal} {\bibinfo  {journal} {J. Chem. Phys.}\ }\textbf {\bibinfo {volume} {9}},\ \bibinfo {pages} {863} (\bibinfo {year} {1941}{\natexlab{a}})}\BibitemShut {NoStop}%
\bibitem [{\citenamefont {Stockmayer}(1941{\natexlab{b}})}]{stockmayer_second_1941-1}%
  \BibitemOpen
  \bibfield  {author} {\bibinfo {author} {\bibfnamefont {W.~H.}\ \bibnamefont {Stockmayer}},\ }\href {\doibase 10.1063/1.1750922} {\bibfield  {journal} {\bibinfo  {journal} {J. Chem. Phys.}\ }\textbf {\bibinfo {volume} {9}},\ \bibinfo {pages} {398} (\bibinfo {year} {1941}{\natexlab{b}})}\BibitemShut {NoStop}%
\bibitem [{\citenamefont {Shock}\ \emph {et~al.}(2020)\citenamefont {Shock}, \citenamefont {Stevens}, \citenamefont {Frischknecht},\ and\ \citenamefont {Nakamura}}]{shock_solvation_2020}%
  \BibitemOpen
  \bibfield  {author} {\bibinfo {author} {\bibfnamefont {C.~J.}\ \bibnamefont {Shock}}, \bibinfo {author} {\bibfnamefont {M.~J.}\ \bibnamefont {Stevens}}, \bibinfo {author} {\bibfnamefont {A.~L.}\ \bibnamefont {Frischknecht}}, \ and\ \bibinfo {author} {\bibfnamefont {I.}~\bibnamefont {Nakamura}},\ }\href {\doibase 10.1021/acs.jpcb.0c00769} {\bibfield  {journal} {\bibinfo  {journal} {J. Phys. Chem. B}\ }\textbf {\bibinfo {volume} {124}},\ \bibinfo {pages} {4598} (\bibinfo {year} {2020})}\BibitemShut {NoStop}%
\bibitem [{\citenamefont {Bagchi}\ and\ \citenamefont {Jana}(2010)}]{bagchi_solvation_2010}%
  \BibitemOpen
  \bibfield  {author} {\bibinfo {author} {\bibfnamefont {B.}~\bibnamefont {Bagchi}}\ and\ \bibinfo {author} {\bibfnamefont {B.}~\bibnamefont {Jana}},\ }\href {\doibase 10.1039/B902048A} {\bibfield  {journal} {\bibinfo  {journal} {Chem. Soc. Rev.}\ }\textbf {\bibinfo {volume} {39}},\ \bibinfo {pages} {1936} (\bibinfo {year} {2010})}\BibitemShut {NoStop}%
\bibitem [{\citenamefont {Perera}\ and\ \citenamefont {Berkowitz}(1992)}]{perera_dynamics_1992}%
  \BibitemOpen
  \bibfield  {author} {\bibinfo {author} {\bibfnamefont {L.}~\bibnamefont {Perera}}\ and\ \bibinfo {author} {\bibfnamefont {M.~L.}\ \bibnamefont {Berkowitz}},\ }\href {\doibase 10.1063/1.461954} {\bibfield  {journal} {\bibinfo  {journal} {J. Chem. Phys.}\ }\textbf {\bibinfo {volume} {96}},\ \bibinfo {pages} {3092} (\bibinfo {year} {1992})}\BibitemShut {NoStop}%
\bibitem [{\citenamefont {Adams}\ and\ \citenamefont {Adams}(1981)}]{adams_static_1981}%
  \BibitemOpen
  \bibfield  {author} {\bibinfo {author} {\bibfnamefont {D.}~\bibnamefont {Adams}}\ and\ \bibinfo {author} {\bibfnamefont {E.}~\bibnamefont {Adams}},\ }\href {\doibase 10.1080/00268978100100701} {\bibfield  {journal} {\bibinfo  {journal} {Molecular Physics}\ }\textbf {\bibinfo {volume} {42}},\ \bibinfo {pages} {907} (\bibinfo {year} {1981})}\BibitemShut {NoStop}%
\bibitem [{\citenamefont {Pollock}\ and\ \citenamefont {Alder}(1980)}]{pollock_static_1980}%
  \BibitemOpen
  \bibfield  {author} {\bibinfo {author} {\bibfnamefont {E.~L.}\ \bibnamefont {Pollock}}\ and\ \bibinfo {author} {\bibfnamefont {B.~J.}\ \bibnamefont {Alder}},\ }\href {\doibase 10.1016/0378-4371(80)90058-8} {\bibfield  {journal} {\bibinfo  {journal} {Physica A: Statistical Mechanics and its Applications}\ }\textbf {\bibinfo {volume} {102}},\ \bibinfo {pages} {1} (\bibinfo {year} {1980})}\BibitemShut {NoStop}%
\bibitem [{\citenamefont {Shock}\ \emph {et~al.}(2023)\citenamefont {Shock}, \citenamefont {Stevens}, \citenamefont {Frischknecht},\ and\ \citenamefont {Nakamura}}]{shock_molecular_2023}%
  \BibitemOpen
  \bibfield  {author} {\bibinfo {author} {\bibfnamefont {C.~J.}\ \bibnamefont {Shock}}, \bibinfo {author} {\bibfnamefont {M.~J.}\ \bibnamefont {Stevens}}, \bibinfo {author} {\bibfnamefont {A.~L.}\ \bibnamefont {Frischknecht}}, \ and\ \bibinfo {author} {\bibfnamefont {I.}~\bibnamefont {Nakamura}},\ }\href {\doibase 10.1063/5.0165481} {\bibfield  {journal} {\bibinfo  {journal} {The Journal of Chemical Physics}\ }\textbf {\bibinfo {volume} {159}},\ \bibinfo {pages} {134507} (\bibinfo {year} {2023})}\BibitemShut {NoStop}%
\bibitem [{\citenamefont {Groh}\ and\ \citenamefont {Dietrich}(1994)}]{groh_ferroelectric_1994}%
  \BibitemOpen
  \bibfield  {author} {\bibinfo {author} {\bibfnamefont {B.}~\bibnamefont {Groh}}\ and\ \bibinfo {author} {\bibfnamefont {S.}~\bibnamefont {Dietrich}},\ }\href {\doibase 10.1103/PhysRevE.50.3814} {\bibfield  {journal} {\bibinfo  {journal} {Phys. Rev. E}\ }\textbf {\bibinfo {volume} {50}},\ \bibinfo {pages} {3814} (\bibinfo {year} {1994})}\BibitemShut {NoStop}%
\bibitem [{\citenamefont {Weis}\ and\ \citenamefont {Levesque}(1993)}]{weis_ferroelectric_1993}%
  \BibitemOpen
  \bibfield  {author} {\bibinfo {author} {\bibfnamefont {J.~J.}\ \bibnamefont {Weis}}\ and\ \bibinfo {author} {\bibfnamefont {D.}~\bibnamefont {Levesque}},\ }\href {\doibase 10.1103/PhysRevE.48.3728} {\bibfield  {journal} {\bibinfo  {journal} {Phys. Rev. E}\ }\textbf {\bibinfo {volume} {48}},\ \bibinfo {pages} {3728} (\bibinfo {year} {1993})}\BibitemShut {NoStop}%
\bibitem [{\citenamefont {Pounds}\ and\ \citenamefont {Madden}(2007)}]{pounds_are_2007}%
  \BibitemOpen
  \bibfield  {author} {\bibinfo {author} {\bibfnamefont {M.~A.}\ \bibnamefont {Pounds}}\ and\ \bibinfo {author} {\bibfnamefont {P.~A.}\ \bibnamefont {Madden}},\ }\href {\doibase 10.1063/1.2672734} {\bibfield  {journal} {\bibinfo  {journal} {J. Chem. Phys.}\ }\textbf {\bibinfo {volume} {126}},\ \bibinfo {pages} {104506} (\bibinfo {year} {2007})}\BibitemShut {NoStop}%
\bibitem [{\citenamefont {Bartke}\ and\ \citenamefont {Hentschke}(2006)}]{bartke_dielectric_2006}%
  \BibitemOpen
  \bibfield  {author} {\bibinfo {author} {\bibfnamefont {J.}~\bibnamefont {Bartke}}\ and\ \bibinfo {author} {\bibfnamefont {R.}~\bibnamefont {Hentschke}},\ }\href {\doibase 10.1080/00268970600961990} {\bibfield  {journal} {\bibinfo  {journal} {Molecular Physics}\ }\textbf {\bibinfo {volume} {104}},\ \bibinfo {pages} {3057} (\bibinfo {year} {2006})}\BibitemShut {NoStop}%
\bibitem [{\citenamefont {Marx}\ \emph {et~al.}(2022)\citenamefont {Marx}, \citenamefont {Kohns},\ and\ \citenamefont {Langenbach}}]{marx_phase_2022}%
  \BibitemOpen
  \bibfield  {author} {\bibinfo {author} {\bibfnamefont {J.}~\bibnamefont {Marx}}, \bibinfo {author} {\bibfnamefont {M.}~\bibnamefont {Kohns}}, \ and\ \bibinfo {author} {\bibfnamefont {K.}~\bibnamefont {Langenbach}},\ }\href {\doibase 10.1002/cite.202255007} {\bibfield  {journal} {\bibinfo  {journal} {Chemie Ingenieur Technik}\ }\textbf {\bibinfo {volume} {94}},\ \bibinfo {pages} {1345} (\bibinfo {year} {2022})}\BibitemShut {NoStop}%
\bibitem [{\citenamefont {Marx}\ \emph {et~al.}(2023)\citenamefont {Marx}, \citenamefont {Kohns},\ and\ \citenamefont {Langenbach}}]{marx_vapor-liquid_2023}%
  \BibitemOpen
  \bibfield  {author} {\bibinfo {author} {\bibfnamefont {J.}~\bibnamefont {Marx}}, \bibinfo {author} {\bibfnamefont {M.}~\bibnamefont {Kohns}}, \ and\ \bibinfo {author} {\bibfnamefont {K.}~\bibnamefont {Langenbach}},\ }\href {\doibase 10.1016/j.fluid.2023.113742} {\bibfield  {journal} {\bibinfo  {journal} {Fluid Phase Equilibria}\ }\textbf {\bibinfo {volume} {568}},\ \bibinfo {pages} {113742} (\bibinfo {year} {2023})}\BibitemShut {NoStop}%
\bibitem [{\citenamefont {Moore}\ \emph {et~al.}(2015)\citenamefont {Moore}, \citenamefont {Stevens},\ and\ \citenamefont {Grest}}]{moore_liquid-vapor_2015}%
  \BibitemOpen
  \bibfield  {author} {\bibinfo {author} {\bibfnamefont {S.~G.}\ \bibnamefont {Moore}}, \bibinfo {author} {\bibfnamefont {M.~J.}\ \bibnamefont {Stevens}}, \ and\ \bibinfo {author} {\bibfnamefont {G.~S.}\ \bibnamefont {Grest}},\ }\href {\doibase 10.1103/PhysRevE.91.022309} {\bibfield  {journal} {\bibinfo  {journal} {Phys. Rev. E}\ }\textbf {\bibinfo {volume} {91}},\ \bibinfo {pages} {022309} (\bibinfo {year} {2015})}\BibitemShut {NoStop}%
\bibitem [{\citenamefont {Frodl}\ and\ \citenamefont {Dietrich}(1992)}]{frodl_bulk_1992}%
  \BibitemOpen
  \bibfield  {author} {\bibinfo {author} {\bibfnamefont {P.}~\bibnamefont {Frodl}}\ and\ \bibinfo {author} {\bibfnamefont {S.}~\bibnamefont {Dietrich}},\ }\href {\doibase 10.1103/PhysRevA.45.7330} {\bibfield  {journal} {\bibinfo  {journal} {Phys. Rev. A}\ }\textbf {\bibinfo {volume} {45}},\ \bibinfo {pages} {7330} (\bibinfo {year} {1992})}\BibitemShut {NoStop}%
\bibitem [{\citenamefont {Frodl}\ and\ \citenamefont {Dietrich}(1993)}]{frodl_thermal_1993}%
  \BibitemOpen
  \bibfield  {author} {\bibinfo {author} {\bibfnamefont {P.}~\bibnamefont {Frodl}}\ and\ \bibinfo {author} {\bibfnamefont {S.}~\bibnamefont {Dietrich}},\ }\href {\doibase 10.1103/PhysRevE.48.3741} {\bibfield  {journal} {\bibinfo  {journal} {Phys. Rev. E}\ }\textbf {\bibinfo {volume} {48}},\ \bibinfo {pages} {3741} (\bibinfo {year} {1993})}\BibitemShut {NoStop}%
\bibitem [{\citenamefont {Iatsevitch}\ and\ \citenamefont {Forstmann}(2000)}]{iatsevitch_structure_2000}%
  \BibitemOpen
  \bibfield  {author} {\bibinfo {author} {\bibfnamefont {S.}~\bibnamefont {Iatsevitch}}\ and\ \bibinfo {author} {\bibfnamefont {F.}~\bibnamefont {Forstmann}},\ }\href {\doibase 10.1080/002689700413569} {\bibfield  {journal} {\bibinfo  {journal} {Molecular Physics}\ }\textbf {\bibinfo {volume} {98}},\ \bibinfo {pages} {1309} (\bibinfo {year} {2000})}\BibitemShut {NoStop}%
\bibitem [{\citenamefont {Mecke}\ \emph {et~al.}(2001)\citenamefont {Mecke}, \citenamefont {Fischer},\ and\ \citenamefont {Winkelmann}}]{mecke_molecular_2001}%
  \BibitemOpen
  \bibfield  {author} {\bibinfo {author} {\bibfnamefont {M.}~\bibnamefont {Mecke}}, \bibinfo {author} {\bibfnamefont {J.}~\bibnamefont {Fischer}}, \ and\ \bibinfo {author} {\bibfnamefont {J.}~\bibnamefont {Winkelmann}},\ }\href {\doibase 10.1063/1.1349177} {\bibfield  {journal} {\bibinfo  {journal} {The Journal of Chemical Physics}\ }\textbf {\bibinfo {volume} {114}},\ \bibinfo {pages} {5842} (\bibinfo {year} {2001})}\BibitemShut {NoStop}%
\bibitem [{\citenamefont {Enders}\ \emph {et~al.}(2004)\citenamefont {Enders}, \citenamefont {Kahl}, \citenamefont {Mecke},\ and\ \citenamefont {Winkelmann}}]{enders_molecular_2004}%
  \BibitemOpen
  \bibfield  {author} {\bibinfo {author} {\bibfnamefont {S.}~\bibnamefont {Enders}}, \bibinfo {author} {\bibfnamefont {H.}~\bibnamefont {Kahl}}, \bibinfo {author} {\bibfnamefont {M.}~\bibnamefont {Mecke}}, \ and\ \bibinfo {author} {\bibfnamefont {J.}~\bibnamefont {Winkelmann}},\ }\href {\doibase 10.1016/j.molliq.2003.12.020} {\bibfield  {journal} {\bibinfo  {journal} {Journal of Molecular Liquids}\ }\textbf {\bibinfo {volume} {115}},\ \bibinfo {pages} {29} (\bibinfo {year} {2004})}\BibitemShut {NoStop}%
\bibitem [{\citenamefont {Eggebrecht}\ \emph {et~al.}(1987{\natexlab{a}})\citenamefont {Eggebrecht}, \citenamefont {Thompson},\ and\ \citenamefont {Gubbins}}]{eggebrecht_liquidvapor_1987}%
  \BibitemOpen
  \bibfield  {author} {\bibinfo {author} {\bibfnamefont {J.}~\bibnamefont {Eggebrecht}}, \bibinfo {author} {\bibfnamefont {S.~M.}\ \bibnamefont {Thompson}}, \ and\ \bibinfo {author} {\bibfnamefont {K.~E.}\ \bibnamefont {Gubbins}},\ }\href {\doibase 10.1063/1.452128} {\bibfield  {journal} {\bibinfo  {journal} {The Journal of Chemical Physics}\ }\textbf {\bibinfo {volume} {86}},\ \bibinfo {pages} {2299} (\bibinfo {year} {1987}{\natexlab{a}})}\BibitemShut {NoStop}%
\bibitem [{\citenamefont {Langenbach}(2017)}]{langenbach_co-oriented_2017}%
  \BibitemOpen
  \bibfield  {author} {\bibinfo {author} {\bibfnamefont {K.}~\bibnamefont {Langenbach}},\ }\href {\doibase 10.1016/j.ces.2017.08.025} {\bibfield  {journal} {\bibinfo  {journal} {Chemical Engineering Science}\ }\textbf {\bibinfo {volume} {174}},\ \bibinfo {pages} {40} (\bibinfo {year} {2017})}\BibitemShut {NoStop}%
\bibitem [{\citenamefont {Kusaka}\ \emph {et~al.}(1995)\citenamefont {Kusaka}, \citenamefont {Wang},\ and\ \citenamefont {Seinfeld}}]{kusaka_ion-induced_1995}%
  \BibitemOpen
  \bibfield  {author} {\bibinfo {author} {\bibfnamefont {I.}~\bibnamefont {Kusaka}}, \bibinfo {author} {\bibfnamefont {Z.-G.}\ \bibnamefont {Wang}}, \ and\ \bibinfo {author} {\bibfnamefont {J.~H.}\ \bibnamefont {Seinfeld}},\ }\href {\doibase 10.1063/1.469158} {\bibfield  {journal} {\bibinfo  {journal} {The Journal of Chemical Physics}\ }\textbf {\bibinfo {volume} {102}},\ \bibinfo {pages} {913} (\bibinfo {year} {1995})}\BibitemShut {NoStop}%
\bibitem [{\citenamefont {Kohns}\ \emph {et~al.}(2020)\citenamefont {Kohns}, \citenamefont {Marx},\ and\ \citenamefont {Langenbach}}]{kohns_relative_2020}%
  \BibitemOpen
  \bibfield  {author} {\bibinfo {author} {\bibfnamefont {M.}~\bibnamefont {Kohns}}, \bibinfo {author} {\bibfnamefont {J.}~\bibnamefont {Marx}}, \ and\ \bibinfo {author} {\bibfnamefont {K.}~\bibnamefont {Langenbach}},\ }\href {\doibase 10.1021/acs.jced.0c00769} {\bibfield  {journal} {\bibinfo  {journal} {J. Chem. Eng. Data}\ }\textbf {\bibinfo {volume} {65}},\ \bibinfo {pages} {5891} (\bibinfo {year} {2020})}\BibitemShut {NoStop}%
\bibitem [{\citenamefont {Stukowski}(2010)}]{ovito}%
  \BibitemOpen
  \bibfield  {author} {\bibinfo {author} {\bibfnamefont {A.}~\bibnamefont {Stukowski}},\ }\href {\doibase {10.1088/0965-0393/18/1/015012}} {\bibfield  {journal} {\bibinfo  {journal} {{MODELLING AND SIMULATION IN MATERIALS SCIENCE AND ENGINEERING}}\ }\textbf {\bibinfo {volume} {{18}}} (\bibinfo {year} {{2010}}),\ {10.1088/0965-0393/18/1/015012}}\BibitemShut {NoStop}%
\bibitem [{\citenamefont {Allen}\ and\ \citenamefont {Tildesley}(2017)}]{allen_computer_2017}%
  \BibitemOpen
  \bibfield  {author} {\bibinfo {author} {\bibfnamefont {M.~P.}\ \bibnamefont {Allen}}\ and\ \bibinfo {author} {\bibfnamefont {D.~J.}\ \bibnamefont {Tildesley}},\ }\href {https://doi.org/10.1093/oso/9780198803195.001.0001} {\emph {\bibinfo {title} {Computer {Simulation} of {Liquids}}}}\ (\bibinfo  {publisher} {Oxford University Press},\ \bibinfo {year} {2017})\BibitemShut {NoStop}%
\bibitem [{\citenamefont {Kamberaj}\ \emph {et~al.}(2005)\citenamefont {Kamberaj}, \citenamefont {Low},\ and\ \citenamefont {Neal}}]{kamberaj_time_2005}%
  \BibitemOpen
  \bibfield  {author} {\bibinfo {author} {\bibfnamefont {H.}~\bibnamefont {Kamberaj}}, \bibinfo {author} {\bibfnamefont {R.~J.}\ \bibnamefont {Low}}, \ and\ \bibinfo {author} {\bibfnamefont {M.~P.}\ \bibnamefont {Neal}},\ }\href {\doibase 10.1063/1.1906216} {\bibfield  {journal} {\bibinfo  {journal} {J. Chem. Phys.}\ }\textbf {\bibinfo {volume} {122}},\ \bibinfo {pages} {224114} (\bibinfo {year} {2005})}\BibitemShut {NoStop}%
\bibitem [{\citenamefont {Toukmaji}\ \emph {et~al.}(2000)\citenamefont {Toukmaji}, \citenamefont {Sagui}, \citenamefont {Board},\ and\ \citenamefont {Darden}}]{toukmaji_efficient_2000}%
  \BibitemOpen
  \bibfield  {author} {\bibinfo {author} {\bibfnamefont {A.}~\bibnamefont {Toukmaji}}, \bibinfo {author} {\bibfnamefont {C.}~\bibnamefont {Sagui}}, \bibinfo {author} {\bibfnamefont {J.}~\bibnamefont {Board}}, \ and\ \bibinfo {author} {\bibfnamefont {T.}~\bibnamefont {Darden}},\ }\href {\doibase 10.1063/1.1324708} {\bibfield  {journal} {\bibinfo  {journal} {J. Chem. Phys.}\ }\textbf {\bibinfo {volume} {113}},\ \bibinfo {pages} {10913} (\bibinfo {year} {2000})}\BibitemShut {NoStop}%
\bibitem [{\citenamefont {Cerdà}\ \emph {et~al.}(2008)\citenamefont {Cerdà}, \citenamefont {Ballenegger}, \citenamefont {Lenz},\ and\ \citenamefont {Holm}}]{cerda_p3m_2008}%
  \BibitemOpen
  \bibfield  {author} {\bibinfo {author} {\bibfnamefont {J.~J.}\ \bibnamefont {Cerdà}}, \bibinfo {author} {\bibfnamefont {V.}~\bibnamefont {Ballenegger}}, \bibinfo {author} {\bibfnamefont {O.}~\bibnamefont {Lenz}}, \ and\ \bibinfo {author} {\bibfnamefont {C.}~\bibnamefont {Holm}},\ }\href {\doibase 10.1063/1.3000389} {\bibfield  {journal} {\bibinfo  {journal} {J. Chem. Phys.}\ }\textbf {\bibinfo {volume} {129}},\ \bibinfo {pages} {234104} (\bibinfo {year} {2008})}\BibitemShut {NoStop}%
\bibitem [{\citenamefont {Jungwirth}\ and\ \citenamefont {Tobias}(2006)}]{jungwirth_specific_2006}%
  \BibitemOpen
  \bibfield  {author} {\bibinfo {author} {\bibfnamefont {P.}~\bibnamefont {Jungwirth}}\ and\ \bibinfo {author} {\bibfnamefont {D.~J.}\ \bibnamefont {Tobias}},\ }\href {\doibase 10.1021/cr0403741} {\bibfield  {journal} {\bibinfo  {journal} {Chem. Rev.}\ }\textbf {\bibinfo {volume} {106}},\ \bibinfo {pages} {1259} (\bibinfo {year} {2006})}\BibitemShut {NoStop}%
\bibitem [{\citenamefont {Matsumoto}\ and\ \citenamefont {Kataoka}(1988)}]{matsumoto_study_1988}%
  \BibitemOpen
  \bibfield  {author} {\bibinfo {author} {\bibfnamefont {M.}~\bibnamefont {Matsumoto}}\ and\ \bibinfo {author} {\bibfnamefont {Y.}~\bibnamefont {Kataoka}},\ }\href {\doibase 10.1063/1.453919} {\bibfield  {journal} {\bibinfo  {journal} {The Journal of Chemical Physics}\ }\textbf {\bibinfo {volume} {88}},\ \bibinfo {pages} {3233} (\bibinfo {year} {1988})}\BibitemShut {NoStop}%
\bibitem [{\citenamefont {Matsumoto}\ and\ \citenamefont {Kataoka}(1989)}]{matsumoto_molecular_1989}%
  \BibitemOpen
  \bibfield  {author} {\bibinfo {author} {\bibfnamefont {M.}~\bibnamefont {Matsumoto}}\ and\ \bibinfo {author} {\bibfnamefont {Y.}~\bibnamefont {Kataoka}},\ }\href {\doibase 10.1063/1.455982} {\bibfield  {journal} {\bibinfo  {journal} {J. Chem. Phys.}\ }\textbf {\bibinfo {volume} {90}},\ \bibinfo {pages} {2398} (\bibinfo {year} {1989})}\BibitemShut {NoStop}%
\bibitem [{\citenamefont {Cendagorta}\ and\ \citenamefont {Ichiye}(2015)}]{cendagorta_surface_2015}%
  \BibitemOpen
  \bibfield  {author} {\bibinfo {author} {\bibfnamefont {J.~R.}\ \bibnamefont {Cendagorta}}\ and\ \bibinfo {author} {\bibfnamefont {T.}~\bibnamefont {Ichiye}},\ }\href {\doibase 10.1021/jp508878v} {\bibfield  {journal} {\bibinfo  {journal} {J. Phys. Chem. B}\ }\textbf {\bibinfo {volume} {119}},\ \bibinfo {pages} {9114} (\bibinfo {year} {2015})}\BibitemShut {NoStop}%
\bibitem [{\citenamefont {Yeh}\ \emph {et~al.}(2001)\citenamefont {Yeh}, \citenamefont {Zhang}, \citenamefont {Held}, \citenamefont {Mebel}, \citenamefont {Wei}, \citenamefont {Lin},\ and\ \citenamefont {Shen}}]{yeh_structure_2001}%
  \BibitemOpen
  \bibfield  {author} {\bibinfo {author} {\bibfnamefont {Y.~L.}\ \bibnamefont {Yeh}}, \bibinfo {author} {\bibfnamefont {C.}~\bibnamefont {Zhang}}, \bibinfo {author} {\bibfnamefont {H.}~\bibnamefont {Held}}, \bibinfo {author} {\bibfnamefont {A.~M.}\ \bibnamefont {Mebel}}, \bibinfo {author} {\bibfnamefont {X.}~\bibnamefont {Wei}}, \bibinfo {author} {\bibfnamefont {S.~H.}\ \bibnamefont {Lin}}, \ and\ \bibinfo {author} {\bibfnamefont {Y.~R.}\ \bibnamefont {Shen}},\ }\href {\doibase 10.1063/1.1333761} {\bibfield  {journal} {\bibinfo  {journal} {The Journal of Chemical Physics}\ }\textbf {\bibinfo {volume} {114}},\ \bibinfo {pages} {1837} (\bibinfo {year} {2001})}\BibitemShut {NoStop}%
\bibitem [{\citenamefont {Martins-Costa}\ and\ \citenamefont {Ruiz-López}(2023)}]{martins-costa_electrostatics_2023}%
  \BibitemOpen
  \bibfield  {author} {\bibinfo {author} {\bibfnamefont {M.~T.~C.}\ \bibnamefont {Martins-Costa}}\ and\ \bibinfo {author} {\bibfnamefont {M.~F.}\ \bibnamefont {Ruiz-López}},\ }\href {\doibase 10.1021/jacs.2c12089} {\bibfield  {journal} {\bibinfo  {journal} {J. Am. Chem. Soc.}\ }\textbf {\bibinfo {volume} {145}},\ \bibinfo {pages} {1400} (\bibinfo {year} {2023})}\BibitemShut {NoStop}%
\bibitem [{\citenamefont {Kuo}\ and\ \citenamefont {Mundy}(2004)}]{kuo_ab_2004}%
  \BibitemOpen
  \bibfield  {author} {\bibinfo {author} {\bibfnamefont {I.-F.~W.}\ \bibnamefont {Kuo}}\ and\ \bibinfo {author} {\bibfnamefont {C.~J.}\ \bibnamefont {Mundy}},\ }\href {\doibase 10.1126/science.1092787} {\bibfield  {journal} {\bibinfo  {journal} {Science}\ }\textbf {\bibinfo {volume} {303}},\ \bibinfo {pages} {658} (\bibinfo {year} {2004})}\BibitemShut {NoStop}%
\bibitem [{\citenamefont {Tobias}\ \emph {et~al.}(2013)\citenamefont {Tobias}, \citenamefont {Stern}, \citenamefont {Baer}, \citenamefont {Levin},\ and\ \citenamefont {Mundy}}]{tobias_simulation_2013}%
  \BibitemOpen
  \bibfield  {author} {\bibinfo {author} {\bibfnamefont {D.~J.}\ \bibnamefont {Tobias}}, \bibinfo {author} {\bibfnamefont {A.~C.}\ \bibnamefont {Stern}}, \bibinfo {author} {\bibfnamefont {M.~D.}\ \bibnamefont {Baer}}, \bibinfo {author} {\bibfnamefont {Y.}~\bibnamefont {Levin}}, \ and\ \bibinfo {author} {\bibfnamefont {C.~J.}\ \bibnamefont {Mundy}},\ }\href {\doibase 10.1146/annurev-physchem-040412-110049} {\bibfield  {journal} {\bibinfo  {journal} {Annual Review of Physical Chemistry}\ }\textbf {\bibinfo {volume} {64}},\ \bibinfo {pages} {339} (\bibinfo {year} {2013})}\BibitemShut {NoStop}%
\bibitem [{\citenamefont {Kathmann}\ \emph {et~al.}(2008)\citenamefont {Kathmann}, \citenamefont {Kuo},\ and\ \citenamefont {Mundy}}]{kathmann_electronic_2008}%
  \BibitemOpen
  \bibfield  {author} {\bibinfo {author} {\bibfnamefont {S.~M.}\ \bibnamefont {Kathmann}}, \bibinfo {author} {\bibfnamefont {I.-F.~W.}\ \bibnamefont {Kuo}}, \ and\ \bibinfo {author} {\bibfnamefont {C.~J.}\ \bibnamefont {Mundy}},\ }\href {\doibase 10.1021/ja802851w} {\bibfield  {journal} {\bibinfo  {journal} {J. Am. Chem. Soc.}\ }\textbf {\bibinfo {volume} {130}},\ \bibinfo {pages} {16556} (\bibinfo {year} {2008})}\BibitemShut {NoStop}%
\bibitem [{\citenamefont {Mundy}\ \emph {et~al.}(2009)\citenamefont {Mundy}, \citenamefont {Kuo}, \citenamefont {Tuckerman}, \citenamefont {Lee},\ and\ \citenamefont {Tobias}}]{mundy_hydroxide_2009}%
  \BibitemOpen
  \bibfield  {author} {\bibinfo {author} {\bibfnamefont {C.~J.}\ \bibnamefont {Mundy}}, \bibinfo {author} {\bibfnamefont {I.-F.~W.}\ \bibnamefont {Kuo}}, \bibinfo {author} {\bibfnamefont {M.~E.}\ \bibnamefont {Tuckerman}}, \bibinfo {author} {\bibfnamefont {H.-S.}\ \bibnamefont {Lee}}, \ and\ \bibinfo {author} {\bibfnamefont {D.~J.}\ \bibnamefont {Tobias}},\ }\href {\doibase 10.1016/j.cplett.2009.09.003} {\bibfield  {journal} {\bibinfo  {journal} {Chemical Physics Letters}\ }\textbf {\bibinfo {volume} {481}},\ \bibinfo {pages} {2} (\bibinfo {year} {2009})}\BibitemShut {NoStop}%
\bibitem [{\citenamefont {Dodia}\ \emph {et~al.}(2019)\citenamefont {Dodia}, \citenamefont {Ohto}, \citenamefont {Imoto},\ and\ \citenamefont {Nagata}}]{dodia_structure_2019}%
  \BibitemOpen
  \bibfield  {author} {\bibinfo {author} {\bibfnamefont {M.}~\bibnamefont {Dodia}}, \bibinfo {author} {\bibfnamefont {T.}~\bibnamefont {Ohto}}, \bibinfo {author} {\bibfnamefont {S.}~\bibnamefont {Imoto}}, \ and\ \bibinfo {author} {\bibfnamefont {Y.}~\bibnamefont {Nagata}},\ }\href {\doibase 10.1021/acs.jctc.9b00253} {\bibfield  {journal} {\bibinfo  {journal} {J. Chem. Theory Comput.}\ }\textbf {\bibinfo {volume} {15}},\ \bibinfo {pages} {3836} (\bibinfo {year} {2019})}\BibitemShut {NoStop}%
\bibitem [{\citenamefont {Horváth}\ \emph {et~al.}(2013)\citenamefont {Horváth}, \citenamefont {Beu}, \citenamefont {Manghi},\ and\ \citenamefont {Palmeri}}]{horvath_vapor-liquid_2013}%
  \BibitemOpen
  \bibfield  {author} {\bibinfo {author} {\bibfnamefont {L.}~\bibnamefont {Horváth}}, \bibinfo {author} {\bibfnamefont {T.}~\bibnamefont {Beu}}, \bibinfo {author} {\bibfnamefont {M.}~\bibnamefont {Manghi}}, \ and\ \bibinfo {author} {\bibfnamefont {J.}~\bibnamefont {Palmeri}},\ }\href {\doibase 10.1063/1.4799938} {\bibfield  {journal} {\bibinfo  {journal} {The Journal of Chemical Physics}\ }\textbf {\bibinfo {volume} {138}},\ \bibinfo {pages} {154702} (\bibinfo {year} {2013})}\BibitemShut {NoStop}%
\bibitem [{\citenamefont {Mecke}\ \emph {et~al.}(1997)\citenamefont {Mecke}, \citenamefont {Winkelmann},\ and\ \citenamefont {Fischer}}]{mecke_molecular_1997}%
  \BibitemOpen
  \bibfield  {author} {\bibinfo {author} {\bibfnamefont {M.}~\bibnamefont {Mecke}}, \bibinfo {author} {\bibfnamefont {J.}~\bibnamefont {Winkelmann}}, \ and\ \bibinfo {author} {\bibfnamefont {J.}~\bibnamefont {Fischer}},\ }\href {\doibase 10.1063/1.475217} {\bibfield  {journal} {\bibinfo  {journal} {J. Chem. Phys.}\ }\textbf {\bibinfo {volume} {107}},\ \bibinfo {pages} {9264} (\bibinfo {year} {1997})}\BibitemShut {NoStop}%
\bibitem [{\citenamefont {Taylor}\ \emph {et~al.}(1996)\citenamefont {Taylor}, \citenamefont {Dang},\ and\ \citenamefont {Garrett}}]{taylor_molecular_1996}%
  \BibitemOpen
  \bibfield  {author} {\bibinfo {author} {\bibfnamefont {R.~S.}\ \bibnamefont {Taylor}}, \bibinfo {author} {\bibfnamefont {L.~X.}\ \bibnamefont {Dang}}, \ and\ \bibinfo {author} {\bibfnamefont {B.~C.}\ \bibnamefont {Garrett}},\ }\href {\doibase 10.1021/jp960615b} {\bibfield  {journal} {\bibinfo  {journal} {J. Phys. Chem.}\ }\textbf {\bibinfo {volume} {100}},\ \bibinfo {pages} {11720} (\bibinfo {year} {1996})}\BibitemShut {NoStop}%
\bibitem [{\citenamefont {Samin}\ \emph {et~al.}(2013)\citenamefont {Samin}, \citenamefont {Tsori},\ and\ \citenamefont {Holm}}]{samin_vapor-liquid_2013}%
  \BibitemOpen
  \bibfield  {author} {\bibinfo {author} {\bibfnamefont {S.}~\bibnamefont {Samin}}, \bibinfo {author} {\bibfnamefont {Y.}~\bibnamefont {Tsori}}, \ and\ \bibinfo {author} {\bibfnamefont {C.}~\bibnamefont {Holm}},\ }\href {\doibase 10.1103/PhysRevE.87.052128} {\bibfield  {journal} {\bibinfo  {journal} {Phys. Rev. E}\ }\textbf {\bibinfo {volume} {87}},\ \bibinfo {pages} {052128} (\bibinfo {year} {2013})}\BibitemShut {NoStop}%
\bibitem [{\citenamefont {Paul}\ and\ \citenamefont {Chandra}(2003)}]{paul_liquidvapor_2003}%
  \BibitemOpen
  \bibfield  {author} {\bibinfo {author} {\bibfnamefont {S.}~\bibnamefont {Paul}}\ and\ \bibinfo {author} {\bibfnamefont {A.}~\bibnamefont {Chandra}},\ }\href {\doibase 10.1021/jp0302820} {\bibfield  {journal} {\bibinfo  {journal} {J. Phys. Chem. B}\ }\textbf {\bibinfo {volume} {107}},\ \bibinfo {pages} {12705} (\bibinfo {year} {2003})}\BibitemShut {NoStop}%
\bibitem [{\citenamefont {Martínez}\ \emph {et~al.}(2009)\citenamefont {Martínez}, \citenamefont {Andrade}, \citenamefont {Birgin},\ and\ \citenamefont {Martínez}}]{martinez_packmol_2009}%
  \BibitemOpen
  \bibfield  {author} {\bibinfo {author} {\bibfnamefont {L.}~\bibnamefont {Martínez}}, \bibinfo {author} {\bibfnamefont {R.}~\bibnamefont {Andrade}}, \bibinfo {author} {\bibfnamefont {E.~G.}\ \bibnamefont {Birgin}}, \ and\ \bibinfo {author} {\bibfnamefont {J.~M.}\ \bibnamefont {Martínez}},\ }\href {\doibase https://doi.org/10.1002/jcc.21224} {\bibfield  {journal} {\bibinfo  {journal} {Journal of Computational Chemistry}\ }\textbf {\bibinfo {volume} {30}},\ \bibinfo {pages} {2157} (\bibinfo {year} {2009})}\BibitemShut {NoStop}%
\bibitem [{\citenamefont {Willard}\ and\ \citenamefont {Chandler}(2010)}]{willard_instantaneous_2010}%
  \BibitemOpen
  \bibfield  {author} {\bibinfo {author} {\bibfnamefont {A.~P.}\ \bibnamefont {Willard}}\ and\ \bibinfo {author} {\bibfnamefont {D.}~\bibnamefont {Chandler}},\ }\href {\doibase 10.1021/jp909219k} {\bibfield  {journal} {\bibinfo  {journal} {J. Phys. Chem. B}\ }\textbf {\bibinfo {volume} {114}},\ \bibinfo {pages} {1954} (\bibinfo {year} {2010})}\BibitemShut {NoStop}%
\bibitem [{\citenamefont {Tarazona}\ \emph {et~al.}(2012)\citenamefont {Tarazona}, \citenamefont {Chacón},\ and\ \citenamefont {Bresme}}]{tarazona_intrinsic_2012}%
  \BibitemOpen
  \bibfield  {author} {\bibinfo {author} {\bibfnamefont {P.}~\bibnamefont {Tarazona}}, \bibinfo {author} {\bibfnamefont {E.}~\bibnamefont {Chacón}}, \ and\ \bibinfo {author} {\bibfnamefont {F.}~\bibnamefont {Bresme}},\ }\href {\doibase 10.1088/0953-8984/24/28/284123} {\bibfield  {journal} {\bibinfo  {journal} {J. Phys.: Condens. Matter}\ }\textbf {\bibinfo {volume} {24}},\ \bibinfo {pages} {284123} (\bibinfo {year} {2012})}\BibitemShut {NoStop}%
\bibitem [{\citenamefont {Ye}\ \emph {et~al.}(2024)\citenamefont {Ye}, \citenamefont {Walker},\ and\ \citenamefont {Wang}}]{ye_mdcraft_2024}%
  \BibitemOpen
  \bibfield  {author} {\bibinfo {author} {\bibfnamefont {B.~B.}\ \bibnamefont {Ye}}, \bibinfo {author} {\bibfnamefont {P.~J.}\ \bibnamefont {Walker}}, \ and\ \bibinfo {author} {\bibfnamefont {Z.-G.}\ \bibnamefont {Wang}},\ }\href {\doibase 10.21105/joss.07013} {\bibfield  {journal} {\bibinfo  {journal} {J. Open Source Softw.}\ }\textbf {\bibinfo {volume} {9}},\ \bibinfo {pages} {7013} (\bibinfo {year} {2024})}\BibitemShut {NoStop}%
\bibitem [{\citenamefont {Johnson}\ \emph {et~al.}(2015)\citenamefont {Johnson}, \citenamefont {Benight}, \citenamefont {Barnes},\ and\ \citenamefont {Robinson}}]{johnson_dielectric_2015}%
  \BibitemOpen
  \bibfield  {author} {\bibinfo {author} {\bibfnamefont {L.~E.}\ \bibnamefont {Johnson}}, \bibinfo {author} {\bibfnamefont {S.~J.}\ \bibnamefont {Benight}}, \bibinfo {author} {\bibfnamefont {R.}~\bibnamefont {Barnes}}, \ and\ \bibinfo {author} {\bibfnamefont {B.~H.}\ \bibnamefont {Robinson}},\ }\href {\doibase 10.1021/acs.jpcb.5b00009} {\bibfield  {journal} {\bibinfo  {journal} {J. Phys. Chem. B}\ }\textbf {\bibinfo {volume} {119}},\ \bibinfo {pages} {5240} (\bibinfo {year} {2015})}\BibitemShut {NoStop}%
\bibitem [{\citenamefont {Eggebrecht}\ \emph {et~al.}(1987{\natexlab{b}})\citenamefont {Eggebrecht}, \citenamefont {Gubbins},\ and\ \citenamefont {Thompson}}]{eggebrecht_liquidvapor_1987-1}%
  \BibitemOpen
  \bibfield  {author} {\bibinfo {author} {\bibfnamefont {J.}~\bibnamefont {Eggebrecht}}, \bibinfo {author} {\bibfnamefont {K.~E.}\ \bibnamefont {Gubbins}}, \ and\ \bibinfo {author} {\bibfnamefont {S.~M.}\ \bibnamefont {Thompson}},\ }\href {\doibase 10.1063/1.452127} {\bibfield  {journal} {\bibinfo  {journal} {The Journal of Chemical Physics}\ }\textbf {\bibinfo {volume} {86}},\ \bibinfo {pages} {2286} (\bibinfo {year} {1987}{\natexlab{b}})}\BibitemShut {NoStop}%
\bibitem [{\citenamefont {Teixeira}\ and\ \citenamefont {Gama}(1991)}]{teixeira_density-functional_1991}%
  \BibitemOpen
  \bibfield  {author} {\bibinfo {author} {\bibfnamefont {P.~I.}\ \bibnamefont {Teixeira}}\ and\ \bibinfo {author} {\bibfnamefont {M.~M. T.~D.}\ \bibnamefont {Gama}},\ }\href {\doibase 10.1088/0953-8984/3/1/009} {\bibfield  {journal} {\bibinfo  {journal} {J. Phys.: Condens. Matter}\ }\textbf {\bibinfo {volume} {3}},\ \bibinfo {pages} {111} (\bibinfo {year} {1991})}\BibitemShut {NoStop}%
\bibitem [{\citenamefont {Teixeira}\ and\ \citenamefont {Gama}(2002)}]{teixeira_orientation_2002}%
  \BibitemOpen
  \bibfield  {author} {\bibinfo {author} {\bibfnamefont {P.~I.~C.}\ \bibnamefont {Teixeira}}\ and\ \bibinfo {author} {\bibfnamefont {M.~M. T.~d.}\ \bibnamefont {Gama}},\ }\href {\doibase 10.1088/0953-8984/14/46/319} {\bibfield  {journal} {\bibinfo  {journal} {J. Phys.: Condens. Matter}\ }\textbf {\bibinfo {volume} {14}},\ \bibinfo {pages} {12159} (\bibinfo {year} {2002})}\BibitemShut {NoStop}%
\bibitem [{\citenamefont {Martins-Costa}\ and\ \citenamefont {Ruiz-Lopez}(2015)}]{martins-costa_solvation_2015}%
  \BibitemOpen
  \bibfield  {author} {\bibinfo {author} {\bibfnamefont {M.~T.~C.}\ \bibnamefont {Martins-Costa}}\ and\ \bibinfo {author} {\bibfnamefont {M.~F.}\ \bibnamefont {Ruiz-Lopez}},\ }\href {\doibase 10.1007/s00214-014-1609-z} {\bibfield  {journal} {\bibinfo  {journal} {Theor Chem Acc}\ }\textbf {\bibinfo {volume} {134}},\ \bibinfo {pages} {17} (\bibinfo {year} {2015})}\BibitemShut {NoStop}%
\bibitem [{\citenamefont {Xiong}\ \emph {et~al.}(2020)\citenamefont {Xiong}, \citenamefont {Lee}, \citenamefont {Zare},\ and\ \citenamefont {Min}}]{xiong_strong_2020}%
  \BibitemOpen
  \bibfield  {author} {\bibinfo {author} {\bibfnamefont {H.}~\bibnamefont {Xiong}}, \bibinfo {author} {\bibfnamefont {J.~K.}\ \bibnamefont {Lee}}, \bibinfo {author} {\bibfnamefont {R.~N.}\ \bibnamefont {Zare}}, \ and\ \bibinfo {author} {\bibfnamefont {W.}~\bibnamefont {Min}},\ }\href {\doibase 10.1021/acs.jpclett.0c02061} {\bibfield  {journal} {\bibinfo  {journal} {J. Phys. Chem. Lett.}\ }\textbf {\bibinfo {volume} {11}},\ \bibinfo {pages} {7423} (\bibinfo {year} {2020})}\BibitemShut {NoStop}%
\end{thebibliography}%

\end{document}

\preprint{AIP/123-QED}

\def\*#1{\mathbf{#1}}
\def\&#1{\mathrm{#1}}

\title{Supplementary Information\textit{ for Stockmayer Fluid with a Shifted Dipole: Interfacial Behavior}}

\author{Samuel Varner}
\altaffiliation{These authors contributed equally.}
\affiliation{Division of Chemistry and Chemical Engineering, California Institute of Technology, Pasadena, CA 91125, United States}

\author{Pierre J. Walker}
\altaffiliation{These authors contributed equally.}
\affiliation{Division of Chemistry and Chemical Engineering, California Institute of Technology, Pasadena, CA 91125, United States}

\author{Ananya Venkatachalam}
\altaffiliation{These authors contributed equally.}
\affiliation{Department of Chemistry, Harvey Mudd College, Claremont, CA 91711, United States}

\author{Bilin Zhuang}
\affiliation{Department of Chemistry, Harvey Mudd College, Claremont, CA 91711, United States}

\author{Zhen-Gang Wang}
\affiliation{Division of Chemistry and Chemical Engineering, California Institute of Technology, Pasadena, CA 91125, United States}
\email{zgw@caltech.edu}

\date{\today}

\begin{abstract}

\end{abstract}

\maketitle

\subsubsection*{Reduced Quantities}
In most cases, the reduced quantities are provided in LAMMPS Lennard-Jones' units.

\begin{center}
\begin{tabular}{||c c c||} 
 \hline
 Name & Symbol & Units \\ [0.6ex] 
 \hline\hline
 dipole moment & $\mu$ & $\sqrt{4\pi\epsilon_0\sigma^3\epsilon}$ \\[0.4ex] 
 \hline
 dipole shift & d & $\sigma$\\[0.4ex] 
 \hline
 density & $\rho$ & $\sigma^{-3}$ \\[0.4ex] 
 \hline
 polarization & $P$ & $\sqrt{\frac{4\pi\epsilon_0\epsilon}{\sigma^3}}$ \\[0.6ex] 
 \hline
 electric field & $E^*_z$ & $\sqrt{\frac{4\pi\epsilon}{\epsilon_0\sigma^3}}$ \\ [0.4ex]
  \hline
 electrostatic potential & $\psi_z^*$ & $\sqrt{\frac{4\pi\epsilon}{\epsilon_0\sigma}}$ \\ [1ex] 
 \hline
\end{tabular}
\end{center}

\subsubsection*{Instantaneous Interfaces}

\begin{figure}[hbt!]
    \centering
    \includegraphics[width=1.0\linewidth]{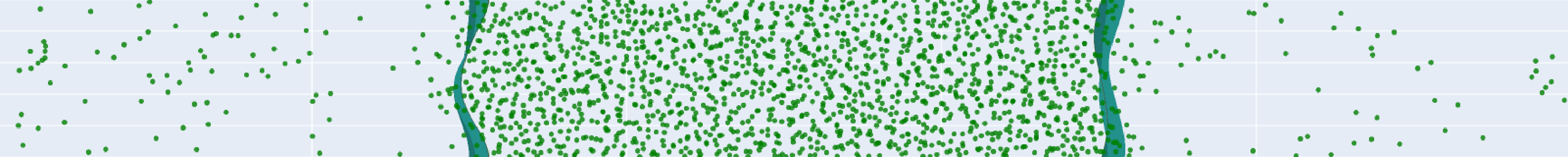}
    \vspace{-1.2cm}
    \caption{Example of instantaneous interfaces computed using the method of Willard and Chandler.\cite{willard_instantaneous_2010}}
    \label{fig:inst-interface-1}
\end{figure}

\begin{figure}[hbt!]
    \centering
    \includegraphics[width=0.6\linewidth]{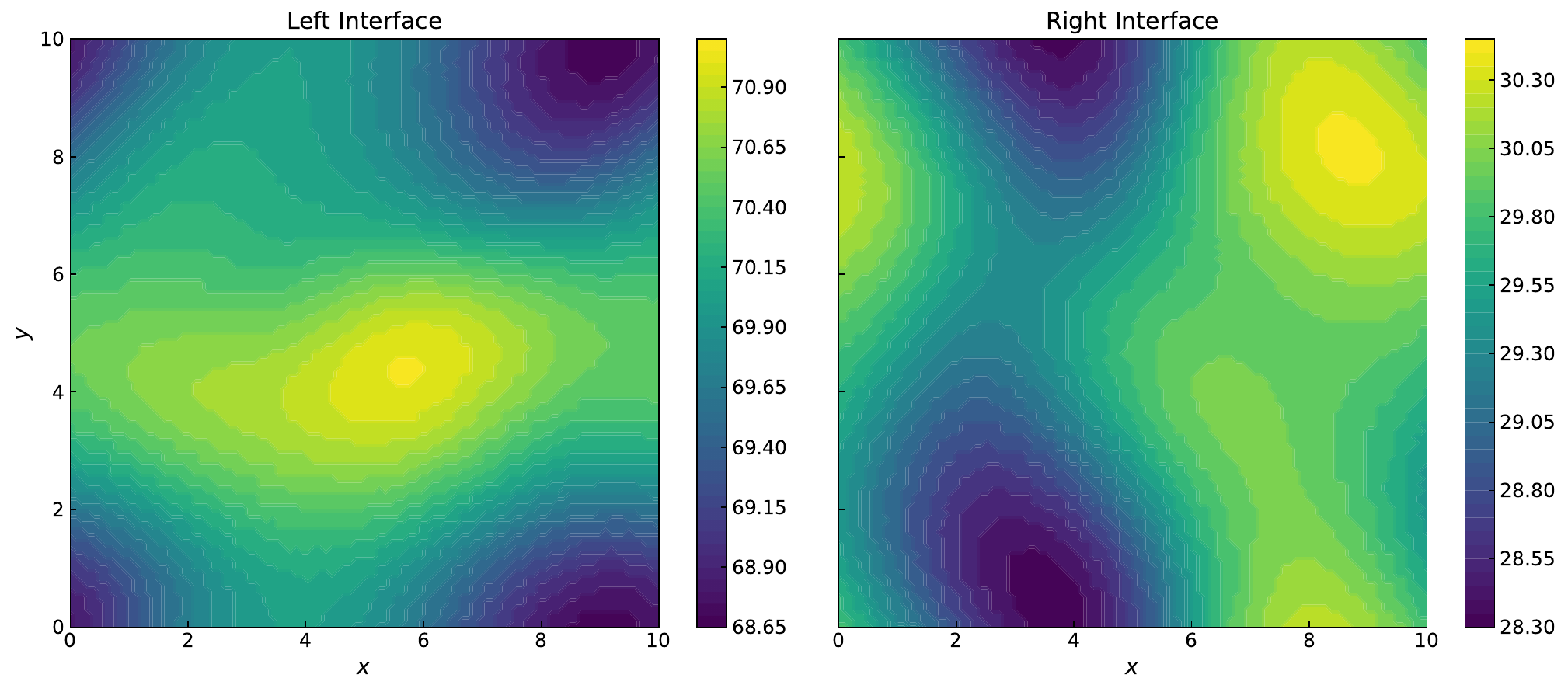}
    \vspace{-0.8cm}
    \caption{2d contour plots of the left and right interfaces from Figure \ref{fig:inst-interface-1}.}
    \label{fig:inst-interface-2d}
\end{figure}

In this work we study the interfacial behavior of polar fluids using a simulation setup containing a slab of liquid in between two vapor regions. To accurately compute profiles of various properties such as density, polarization, electric field, and electrostatic potential, we require a consistent and rigorous definition of the interface position. We elect to use the definition of the instantaneous interface developed by Willard and Chandler.\cite{willard_instantaneous_2010} In Figures \ref{fig:inst-interface-1} and \ref{fig:inst-interface-2d} we provide examples of the instantaneous interfaces computed from a single simulation snapshot.


\clearpage
\subsubsection*{Electric Fields and Electrostatic Potentials}
\begin{figure}[hbt!]
    \centering
    \includegraphics[width=1.0\linewidth]{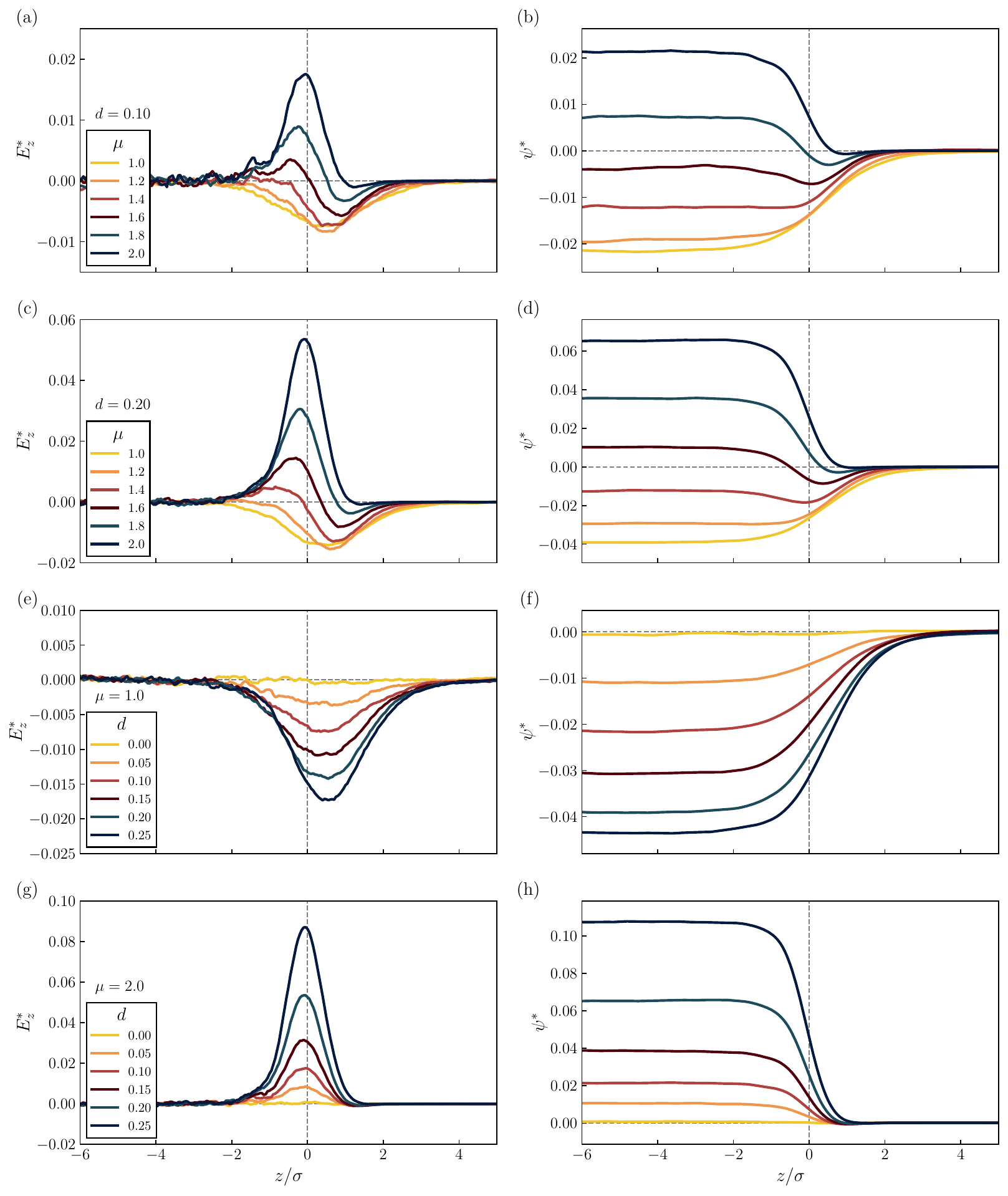}
    \caption{Electric field and electrostatic potential for various combinations of $\mu$ and $d$.}
    \label{fig:E_psi_z_}
\end{figure}

\clearpage
\bibliography{references,extra,ovito-ref}